\documentclass[11pt,notitlepage,tightenlines,eqsecnum,floats,aps,amsmath,superscriptaddress,amssymb,nofootinbib,longbibliography]{revtex4-2}

\usepackage{graphicx}
\usepackage{amsmath,amssymb}
\usepackage{hyperref}
\usepackage[dvipsnames]{color,xcolor,colortbl}
\usepackage[normalem]{ulem}
\usepackage{comment}
\usepackage{wrapfig}
\usepackage{braket}
\usepackage{mathrsfs}
\usepackage[scr=boondoxo]{mathalpha}
\usepackage{xcolor}
\usepackage{leftindex}
\usepackage{float}

\usepackage{color}

\begin{document}

\title{Quasinormal modes in effective loop quantum gravity and consequences for isospectrality}

\author{Alejandro García-Quismondo}
\email{agarciaquismondo@comillas.edu}
\affiliation{ICAI, Universidad Pontificia Comillas, C/ Alberto Aguilera 25, 28015 Madrid, Spain.}
\author{Guillermo A. Mena Marugán}
\email{mena@iem.cfmac.csic.es}
\affiliation{Instituto de Estructura de la Materia, IEM-CSIC, C/ Serrano 121, 28006 Madrid, Spain.}
 \author{Alvaro Torres-Caballeros}
\email{alvaro.torres@iem.cfmac.csic.es}
\affiliation{Instituto de Estructura de la Materia, IEM-CSIC, C/ Serrano 121, 28006 Madrid, Spain.}

\date{August 2026}

\begin{abstract}
We investigate the quasinormal mode spectra of axial and polar perturbations about the effective Ashtekar, Olmedo, and Singh black hole geometry within the hybrid approach to loop quantum gravity. We also compare our results with those previously obtained by del-Corral and Olmedo for the dressed metric approach. Starting from mode equations that are straightforward effective counterparts of the classical equations, we compute these spectra by means of a high-order WKB method with Padé resummation. The violation of isospectrality between axial and polar perturbations that was claimed to exist in the dressed metric approach persists with the hybrid quantization, with deviations of the same order of magnitude as in the former effective case, although slightly larger for the hybrid prescription. In both situations, the isospectrality breaking decreases with the cubic root of the squared black hole mass in Planck units, consistently recovering the classical Schwarzschild limit. In addition, we perform a careful assessment of the applicability of the WKB approximation to the present effective analysis, confirming the parameter regions where the method remains reliable and highlighting specific mode cases for which the approximation breaks down.
\end{abstract}

\maketitle

\section{Introduction}
The pursuit of observational signatures of quantum gravity is arguably one of the most challenging yet pivotal objectives in modern physics. Detecting such signatures is essential for bringing theoretical frameworks of quantum gravity into the realm of falsifiable science, and thus establishing meaningful observational constraints on candidate models. One of the principal difficulties lies in the fact that such effects are expected to become significant only near the Planck scale, where the classical notion of spacetime is expected to break down owing to quantum effects, and which seems extraordinarily difficult to access observationally or experimentally. Physical scenarios such as the pre-inflationary cosmic microwave background or black hole mergers may offer valuable opportunities to probe strong-gravity regimes. Within a merger event, the ringdown phase is particularly attractive because it can be modeled to a large extent using analytical perturbation theory, while the corresponding quasinormal modes that describe the gravitational radiation during this phase can be computed through semianalytical or numerical methods. Proposed next-generation detectors such as LISA \cite{Lisaone} or the Einstein Telescope \cite{SET} are expected to enable high-precision observations of black hole mergers, providing a promising avenue to test general relativity (GR) \cite{Lisatwo,LISAthree,ETtwo,Tests,Berti} and opening opportunities to eventually study deviations predicted by other theories, including alternatives with quantum corrections.

Quasinormal modes represent the natural oscillations of a black hole as it settles back into a stable configuration. This relaxation process, following a fast transition period, is characterized by the dissipation of energy to spatial infinity and the flux of energy across the event horizon, giving rise to inherently damped modes \cite{Nollert}. Beyond describing these characteristic frequencies and damping rates, quasinormal modes may also provide a valuable window into quantum gravity effects. According to the no-hair theorem of classical GR, the quasinormal modes of an uncharged black hole must depend strictly on its mass and spin. Historically, Regge and Wheeler first inspected the axial perturbations \cite{ReggeWheeler}, while Zerilli later analyzed the polar sector \cite{Zerilli}. Chandrasekhar and Detweiler subsequently showed that, for this Schwarzschild case, the axial and polar perturbations are isospectral \cite{Chandra}, i.e., despite being governed by different master equations owing to their distinct potentials, both sectors lead to the exact same quasinormal mode frequencies \cite{Yurov,Glampe}. However, effective quantum modifications introduce the possibility of observing deviations from the classical ringdown spectrum that are parity dependent, therefore resulting in a violation of isospectrality. Hence, isospectrality constitutes a valuable tool to probe potentially new physics.

One of the leading approaches to formulate a quantum theory of gravity is loop quantum gravity (LQG), which is based on a non-perturbative, diffeomorphism-invariant, and background-independent quantization of GR \cite{AP,AB}. In its Hamiltonian formulation, the canonical variables are the Ashtekar-Barbero connection and its conjugate densitized triad, both carrying SU(2) gauge structure and encoding the gravitational degrees of freedom. Rather than promoting the connection directly to an operator, the quantum theory is formulated in terms of holonomies of the connection and fluxes of the triad, leading to a polymer-like quantization with discrete spectra for geometric operators, which results in the existence of a smallest admissible, non-zero eigenvalue for the area operator, known as the area gap \cite{AshLewa, Thiemann}. The application of LQG techniques to symmetry-reduced models has led to the birth of the field of loop quantum cosmology (LQC). This framework has attracted considerable attention owing to its successful resolution of classical big bang singularities in cosmological models, replacing them with a quantum-geometric phenomenon known as the big bounce \cite{APSLetter,APS,MMO}, while also giving rise to potentially testable imprints of pre-inflationary quantum geometry in the large-scale anisotropies of the cosmic microwave background \cite{largescale,OrigAnali,Antonio}.

Motivated by the success of LQC in cosmological setups, similar techniques have been applied to black hole spacetimes. This explains the increasing attention that the study of LQC corrections to the spectra of quasinormal modes has attracted in recent years (for a non-exhaustive list of works on the subject, see e.g. Refs. \cite{CorralOlmedo,Bouh,QNMloop,QNMloop1,QNMloop2,QNMloop3}). 
In fact, since the interior region of a non-rotating black hole can be described as a homogeneous but anisotropic spacetime, namely a Kantowski–Sachs cosmology, the quantization methods of LQC can be suitably imported into the black hole context, where the gravitational phase space is typically parametrized by two canonical pairs, associated with the angular and radial parts of the geometry. Among the effective models of black holes in LQC (see e.g. Refs. \cite{LitBH1,LitBH2,LitBH3,LitBH4,CorichiSingh,AOS,AOS2,Properties}), the one put forward by Ashtekar, Olmedo, and Singh (AOS) constitutes one of the most appealing proposals, in which the classical black hole singularity is resolved and replaced by a transition surface separating trapped and anti-trapped regions of spacetime, often interpreted heuristically as a black-to-white-hole interior transition. The model also admits a well-defined Arnowitt-Deser-Misner (ADM) energy and predicts a correction to the Hawking temperature \cite{AOS,AOS2,Properties}. Moreover, an important achievement of the AOS framework is that the effective spacetime can be consistently extended beyond the black hole interior through a complex canonical transformation, allowing the construction of an effective geometry including the exterior region \cite{AOS, AOS2}. More recently, attention has turned to the application of this complex symplectomorphism to test fields taking into account perturbative degrees of freedom and also to perturbations of the geometry itself \cite{Axial,Polar}. The analysis has been made possible by rewriting the dynamics of the gauge-invariant perturbations in terms of phase space variables, so the employment of the transformation is naturally applicable \cite{GnR,Beatles}. 

However, there is a nuance regarding the precise prescription for implementing effective quantum corrections. In the existing literature, two main approaches that do not affect the ultraviolet behavior of the perturbations have been widely discussed: the dressed metric approach \cite{dressed1,dressed2,dressed3,dressed4} and the hybrid approach \cite{hybrid1,hybrid2,hybrid5,EliMenRev}. On the one hand, the dressed metric approach starts by perturbing the effective, quantum-corrected line element directly. By performing a perturbative treatment to this setup, \textit{dressed} with quantum corrections, one derives the corresponding effective axial and polar potentials \cite{CorralOlmedo}. On the other hand, the hybrid approach begins by perturbing the \textit{naked} classical line element, expressed in terms of the unregularized metric functions \cite{Axial,Polar}. Within this framework, following a perturbative Hamiltonian treatment in the presence of constraints, a generalized version of the master equation is derived \cite{Masterfun,BlackPumas}, which incorporates corrections in the quantum system. In practice, assuming sharply peaked background states in our model, it seems in principle reasonable to implement these corrections by evaluating the canonical, phase space variables within the axial and polar potentials along the effective AOS trajectories. Crucially, the two described schemes are \textit{not} equivalent.\footnote{Even when evaluated on effective background geometries, the dynamics generated by the Hamiltonian constraint of the perturbed canonical system is not the same at all derivative orders as the dynamics ruled by the effective background Hamiltonian which incorporates the quantum modifications of the dressed metric.} 

The dressed metric approach was employed in Ref. \cite{CorralOlmedo} to investigate the quasinormal modes of the AOS model and test whether isospectrality holds after the inclusion of quantum effects. Within that framework, deviations from GR were found to emerge at the fourth significant digit for black holes with a mass $M$ of $500$ Planck units. Furthermore, the authors reported that isospectrality is broken, although the extent of this violation varies as $(2M)^{-2/3}$. Whether these results depend on the quantization prescription employed in LQG/LQC for the perturbations remains an open question. 

In this regard, the main objective of the present work is to analyze whether the violation of isospectrality is sensitive to the chosen quantization scheme within the context of the effective AOS model. For this purpose, and as a first approach, we investigate the quasinormal mode spectrum opting for a completely straightforward (and, thus, naive) implementation of the effective regime for the hybrid quantization proposal. More concretely, as we have already commented, we evaluate the derived potentials in the mode equations directly along the effective AOS trajectories, without introducing intermediate or refined modifications during their construction. By employing the same semianalytical WKB method as Ref. \cite{CorralOlmedo} to compute the quasinormal modes, we are perfectly poised to make a direct and systematic comparison with the dressed metric results reported in that reference and qualitatively determine whether the violation of isospectrality also arises within the hybrid scenario.

Furthermore, we also take the opportunity to sharply inspect the range of validity of the semianalytical WKB method employed in the computation of the quasinormal mode spectrum. Although the WKB approximation has been widely used in the literature and provides reliable results within certain regimes, a further examination of its accuracy and limitations in the context of the potentials of the effective AOS model remains largely unexplored. In particular, the validity of the method is known to depend on the relation between the overtone number $n$ and the angular momentum multipole $l$, with the approximation generally becoming more accurate in the so-called eikonal regime, where $l \gg n$ \cite{KZZ}. While a fully exhaustive analysis lies beyond the scope of this work, we nevertheless aim to provide a more careful assessment of the reliability of the computed spectra for the different regions of the parameter space. Such an analysis will allow us to better characterize the stability of the chosen order in the WKB approximation and the robustness of the claims about the loss of isospectrality, concerning the comparison between the dressed metric and hybrid quantization prescriptions.

The remainder of this article is organized as follows. In Sec. \ref{sec2}, we introduce the Kantowski-Sachs spacetimes in their exterior-region extended form. We then review the effective AOS model, recalling the associated regularization (or polymerization) parameters and the resulting effective solutions to the equations of motion. In Sec. \ref{sec3}, we review the notion of quasinormal modes, the corresponding master equations, and the open boundary conditions that characterize these modes. Within this section, we also discuss the dressed metric (Sec. \ref{subsec3a}) and hybrid (Sec. \ref{subsec3b}) prescriptions for the quantum treatment of the perturbations, and present the four effective perturbative potentials considered in this work (namely the axial and polar potentials associated with each prescription). In Sec. \ref{sec4}, for completeness and self-consistency, we briefly review the semianalytical WKB method, while referring to the relevant literature for more detailed discussions. Subsequently, we present the computed quasinormal mode spectra for the Schwarzschild case and for the AOS effective model, within both the dressed metric and hybrid approaches. We consider the overtone range $0 \leq n \leq 1$ and multipoles $2 \leq l \leq 4$. We also present several plots illustrating the deviations of the quasinormal mode frequencies for the effective background from their Schwarzschild counterparts, as well as the absolute deviations between the results for the two prescriptions incorporating quantum geometry effects on the perturbations. In Sec. \ref{sec5}, we further investigate the regime of applicability of the WKB method and discuss cases identified as \textit{critical}, for which our semianalytical computation approaches its limits of validity. Finally, Sec. \ref{sec6} contains the conclusions of our study and some additional remarks. Throughout our discussion, we employ Planck units, setting $c=G=\hbar=1$.

\section{Kantowski-Sachs spacetimes and effective AOS model\label{sec2}}

Let us start by considering a symmetry-reduced spacetime foliated by a family of homogeneous and anisotropic spacelike surfaces, which is isomorphic to the family of Kantowski-Sachs cosmologies describing the interior of a non-rotating black hole. The line element describing the exterior geometry of the black hole is taken to be an extension of its interior counterpart, adopting the form \cite{AOS2,BlackPumas}
\begin{equation}\label{line}
    ds^2 = -\frac{p_b^2}{L_{0}^2|p_c|}~dt^2+\frac{p_b^2|p_c|}{L_{0}^2}\tilde{N}^2~dx^2+|p_c|~ d\Omega_2^2~,
\end{equation}
where $d\Omega_2^2$ is the line element of the 2-sphere, the coordinate $x$ can be related to the Schwarzschild radial coordinate by a suitable choice of the metric functions, $L_{0}$ is a compactification period for this $x$-coordinate that serves as an infrared regulator, and $\tilde{N}=(4\pi L_{0}) \mathbf{N}/\mathcal{V}$ is the densitized {\it{lapse}} that results from densitizing the lapse function $\mathbf{N}$ over the volume $\mathcal{V}$ of the sections of constant $x$ (i.e., the spatial sections in the interior region). 

The extension from the interior geometry to the exterior form above can be performed by the complex symplectomorphism\footnote{The extension adopted in Ref.~\cite{AOS2} differs from the convention used here by a sign in the configuration variable of the $b$-sector. However, both conventions lead to equivalent line elements and dynamics.} $b\to -i\tilde{b}$, $p_b\to i \tilde{p}_b$, $c\to\tilde{c}$, and $p_c \to \tilde{p}_c$. For simplicity, we drop the tilde notation after implementing the mapping. On the other hand, it is understood that the old and new phase space variables of the model are real in the interior and exterior regions, respectively. Also, note that the timelike and spacelike characters of the coordinates $t$ and $x$ are interchanged when performing the aforementioned extension. 

The metric variables  $p_b(x)$ and $p_c(x)$ are directly related to densitized triads and are canonically conjugate to the variables $b(x)$ and $c(x)$, respectively, which characterize the Ashtekar-Barbero connection. As emphasized by our notation, these variables are functions solely of the $x$-coordinate. They satisfy the canonical commutation relations $\{b,p_b\}=\gamma$
and $\{c,p_c\}=2\gamma$, where $\gamma=0.2375$ is the Immirzi parameter \cite{Immirzi}.

\subsection{Effective AOS dynamics\label{Subsec2a}}

The effective AOS solutions incorporate quantum corrections through two regularization (or polymerization) parameters $\delta_b$ and $\delta_c$, which characterize the coordinate lengths of the holonomy loops used in the LQG description of the system. In order to obtain a good semiclassical behavior, the so-called improved dynamics prescription suggests the physical area of the elementary loop should equal the minimum area eigenvalue. However, to implement this prescription in the AOS model, the physical area of the elementary holonomy loop is evaluated on the transition surface of the effective metric. As a result, the polymerization parameters are not (universal) constants, but rather Dirac observables, in the sense that they remain constant along each given effective trajectory, although they can vary from one effective dynamical trajectory to another \cite{AOS2,Articulin}. In fact, as discussed in Refs. \cite{AOS,AOS2}, this prescription uniquely determines the explicit asymptotic dependence of the parameters on the black hole mass \cite{AOS2}:
\begin{equation}
    \delta_b = \left(\frac{\sqrt{\Delta}}{\sqrt{2\pi}\gamma^2M} \right)^{1/3} \;\;\; \text{and} \;\;\; \; L_{0}\delta_c=\frac{1}{2}\left(\frac{\gamma\Delta^2}{4\pi^2M}\right)^{1/3},
\end{equation}
where $\Delta=5.17$ refers to the area gap (in Planck units and for the value of the Immirzi parameter that we have specified above), i.e., the minimum non-zero eigenvalue allowed for the area operator.

By introducing the lapse function 
\begin{equation}\label{lapse}
    \mathbf{N}=\frac{\gamma\,\delta_b~\text{sgn}(p_c)\,|p_c|^{1/2}}{\sinh(\delta_b b)},
\end{equation}
which is the effective counterpart of the classical lapse $\mathbf{N}_{\text{cl}}=\gamma b^{-1}\text{sgn}(p_c)|p_c|^{1/2}$, with $\text{sgn}(\cdot )$ denoting the sign function, it is possible to find a completely analytical set of solutions to the equations of motion. For the Hamiltonian constraint 
\begin{equation}
\mathscr{H}[\mathbf{N}]= -\dfrac{L_{0}}{2 \Omega_b} \left(\Omega_b^2+ 2\Omega_b\Omega_c -\dfrac{p_b^2}{L_{0}^2}\right),\label{hamconstraint}\end{equation}
where
\begin{equation}\Omega_b=\dfrac{\sinh (\delta_b b) p_b}{\gamma L_{0} \delta_b},\qquad\Omega_c=\dfrac{\sin (\delta_c c) p_c}{\gamma L_{0} \delta_c},\label{Omegas}
\end{equation}
the dynamical solutions can be shown to take the form\footnote{We choose the positive branch of $\tan(\delta_c c/2)$ so that the sign of the constant of motion $\bar{\Omega}_c$ matches the chosen sign convention of the mass $M$.}
\begin{equation}\label{eom}
  \begin{array}{cc}
  \cosh(\delta_bb)=\varepsilon_b \tanh\left( \dfrac{\varepsilon_b X}{2}+C_2\right),   &  \tan\left(\dfrac{\delta_c c}{2}\right)=\mp C_1 e^{-2X}, \\
  \dfrac{p_b}{L_{0}}=-2\dfrac{\delta_b}{\delta_c}\dfrac{\sinh(\delta_b b)}{\gamma^2\delta_b^2-\sinh^2(\delta_b b)}  \dfrac{p_c\sin(\delta_c c)}{L_{0}},& \qquad \,p_c = 4M^2(e^{2X}+C_1^2e^{-2X}),
\end{array}  
\end{equation}
where $X$ is the specific $x$-coordinate associated with the choice of lapse $\mathbf{N}$ and $\varepsilon_b= \sqrt{1+\gamma^2 \delta_b^2}$. We choose the integration constants $C_1= \gamma L_{0} \delta_c/(8M)$ and $C_2= \tanh^{-1}(1/\varepsilon_b)$ to match the Schwarzschild metric (once the respective coordinate is introduced) in the classical limit \cite{AOS2,CorichiSingh}. 

Let us comment that, by inspecting the classical Hamiltonian (therefore using $\mathbf{N}_{\text{cl}}$, for example, in Ref. \cite{AOS2} or equivalently by recalling the classical solutions presented in the same reference), we can straightforwardly see that the quantity $\bar{\Omega}_c= cp_c/(L_{0}\gamma)$ is a Dirac observable, i.e., a constant of motion that, with the above choice of integration constants, is identified with the ADM mass of the black hole. This identification clearly continues to hold when evaluating it on AOS effective trajectories, as can be directly checked from Eq. \eqref{hamconstraint} with the commutation relations given below Eq. \eqref{line}. Indeed, we have
\begin{equation}
    \Omega_c= \frac{\sin(\delta_cc)p_c}{\gamma L_{0} \delta_c} = |M|\,\text{sgn}(M) \,\frac{e^{-2X}}{1+C_1^2e^{-4X}}\left(e^{2X}+C_1^2e^{-2X}\right) = |M| \,\text{sgn}(M),
\end{equation}
where we have employed the identity $\sin(\delta_c c)=2\tan(\delta_c c/2)/[1+\tan^2(\delta_c c/2)]$. Using this identification we can simplify $p_b/L_{0}$ in Eq. \eqref{eom} and write
\begin{equation}\label{simplify}
    \frac{p_b}{L_{0}}=-2|M|\, \text{sgn}(M)\frac{\sinh(\delta_b b)}{\gamma^2\delta_b^2-\sinh^2(\delta_bb)}\sqrt{\varepsilon_b^2-1}.
\end{equation}

\section{\label{sec3} Potentials with quantum corrections and quasinormal modes}

After perturbing the metric and decomposing the perturbations into tensor spherical harmonics, the modes can be separated into axial and polar depending on how they behave under parity transformations. Regge and Wheeler showed that the axial perturbations of a static, spherically symmetric spacetime obey a single master equation \cite{ReggeWheeler}, while Zerilli obtained the analogous result for the polar sector \cite{Zerilli}. Owing to the staticity and spherical symmetry of the background spacetime, the time dependence can be separated through a Fourier decomposition ($e^{-i\tilde{\omega} t}$, with $\tilde{\omega}$ denoting a \textit{dimensionful} frequency), reducing the perturbation dynamics in both parity sectors to a second-order (radial-like in the exterior) differential equation of the form

\begin{equation}\label{QNMs}
    \frac{\partial^2\psi_{\tilde{\omega},l}}{\partial r_{\star}^2} + \left[ \tilde{\omega}^2-V_{l}(r)\right] \psi_{\tilde{\omega},l}=0,
\end{equation}
where $r_{\star}$ is the tortoise coordinate defined by 
\begin{equation}
    dr_{\star}=|p_c|N_0 \,dx,
\end{equation}
and is equal to $r_{\star}=r+2M \ln[r/(2M)-1]$ for the Schwarzschild solution. The potential $V_{l}$ in Eq. \eqref{QNMs} is different for the axial and polar sectors. Chandrasekhar found a transformation 
which classically connects the axial and polar potentials \cite{Chandra2} and has traditionally been regarded as a hidden symmetry of black hole perturbations. However, employing a gauge-invariant Hamiltonian formulation for these perturbations, it has recently been understood that this Chandrasekhar transformation can in fact be seen as a background-dependent canonical transformation \cite{GA2}. Moreover, this transformation is a particular example of the so-called family of Darboux transformations \cite{Darboux,CM1,CM2} and its existence guarantees that the axial and polar sectors of Schwarzschild perturbations are isospectral. In the following, we investigate whether the potentials obtained in LQG through a naive, direct evaluation on the AOS effective trajectories preserve this isospectrality.

Quasinormal modes are solutions of Eq. \eqref{QNMs} with complex Fourier frequency, characterized by the following boundary conditions:
\begin{equation}
    \begin{array}{ccc}
        \psi_{n,l}(r) \propto e^{i\,\tilde{\omega}_{n,l} \,r_{\star}}, & r_{\star}\to-\infty & \text{(horizon),} \\
        \psi_{n,l}(r) \propto e^{-i\,\tilde{\omega}_{n,l} \,r_{\star}}, & r_{\star}\to+\infty & \text{(spatial infinity)}.
    \end{array}
\end{equation}
Note the indices introduced into our former notation for the Fourier frequency $\tilde{\omega}$. The multipole label $l$ appears because each master equation depends on it through its potential (and, therefore, so do its solutions), while the overtone index $n$ labels the countable set of actual quasinormal mode frequencies. Regarding the boundary conditions, the first is a natural physical requirement stating that no radiation emerges from the black hole horizon. Hence, we select purely ingoing waves at this horizon (identified using null coordinates). The second boundary condition imposes that no incoming radiation comes from spatial infinity \cite{Frolov}. This reflects the fact that the system is not externally driven, but rather describes the response of a black hole to a perturbation, during which energy is carried away. This leakage makes the black hole an open dissipative system and explains why perturbations decay in time, allowing the spacetime to relax back to its stationary configuration. 

Since, in the frequency-domain formulation and for the exterior region, Eq. \eqref{eom} does not correspond to a self-adjoint differential operator under the above boundary conditions, it does not amount to a Sturm--Liouville problem in the standard sense. A quasinormal mode is then determined by a value of the complex frequency $\tilde{\omega}_{n,l}$ for which the solution satisfies those boundary conditions. Note that, in general, solutions adapted only to either the condition on the horizon or the condition at infinity are linearly independent \cite{Nollert}, and it is only for quasinormal modes that they become the same. Therefore, when the Green function associated to the perturbations is analytically continued into the complex plane, its poles, located at the zeros of the Wronskian of the mode solutions, turn out to characterize the quasinormal mode frequencies. The resulting spectrum of quasinormal modes $\tilde{\omega}_{n,l}$ consists of complex frequencies with negative imaginary part (in the time-dependence convention $e^{-i\tilde{\omega} t}$), ensuring temporal decay (mode stability). 

Finally, let us comment that in the quasinormal mode analysis presented in Sec. \ref{sec4}, we assume that the boundary conditions are preserved in the effective description. This assumption seems reasonable since describing quasinormal modes via complex poles of the Green function can be seen as a generalization of their standard characterization in terms of boundary conditions.\footnote{A more detailed discussion of these points will be the subject of future research.}

In the following subsections, we review the two main approaches adopted in the literature to incorporate the influence of the (effective) AOS quantum geometry on the perturbations ---the dressed metric and hybrid approaches--- and present the precise form of the potentials that govern the corresponding dynamics of axial and polar perturbations.

\subsection{Axial and polar potentials for the effective AOS model in the dressed metric approach}\label{subsec3a}

One alternative to include LQC corrections affecting the background into the dynamics of the perturbations is the so-called dressed metric approach \cite{dressed1,dressed2,dressed3,dressed4}. For the purposes of the present analysis, this approach consists in evaluating the line element presented in Eq. \eqref{line}, along with the lapse function of Eq. \eqref{lapse}, on the effective solutions introduced in Eqs. \eqref{eom} and \eqref{simplify}. We thus obtain an effective metric incorporating (or dressed with) quantum gravity corrections. By introducing the parameter $\epsilon=1-\varepsilon_b$, the resulting line element can be expressed as 
\begin{equation}
     ds_{\text{eff}}^2= -G_{\text{AOS}}(r) \,dt^2 + F_{\text{AOS}}(r) \,dr^2+ H_{\text{AOS}}(r) \,d\Omega^2,
\end{equation}
with 

\begin{equation}\label{metricAOS}
    \begin{aligned}
            G_{\text{AOS}}&=\frac{r_s^2 (\chi-1) (\mathscr{S}^2\chi-\epsilon^2)(\mathscr{S}\chi+\epsilon)^2}{16(1+\epsilon)^4\chi^2} \, \left[r^{2}\left( 1+ \frac{\gamma^2  L_{0}^2{\delta}_c^2\, r_s^2}{16r^4}\right)\right]^{-1} ,\\
            F_{\text{AOS}}&= \frac{(\mathscr{S}\chi+\epsilon)^2}{(\chi-1)(\mathscr{S}^2\chi-\epsilon^2)}\, \left[r^2 \left( 1+ \frac{\gamma^2   L_{0}^2 {\delta}_c^2 \, r_s^2}{16r^4}\right)\right] ,\\
            H_{\text{AOS}} &= r^2 \left(1+ \frac{\gamma^2   L_{0}^2{\delta}_c^2 \,r_s^2}{16r^4}\right),
    \end{aligned}
\end{equation}
where we have introduced the Schwarzschild radius $r_s=2M$ and the {Schwarzschild coordinate}\footnote{Strictly speaking, $r$ may be thought of as an effective Schwarzschild coordinate, since the lapse function employed here is different from $\mathbf{N}_{\text{cl}}$.} $r$, which satisfies $r>r_s$ and is related to our previous radial-like coordinate $X$ by $r=r_s \, e^{X}$.
In the previous equation, we have also defined the quantities  
\begin{equation}\label{convenience}
\mathscr{S}=2+\epsilon,\qquad R=\frac{r}{r_s},\qquad \chi=R^{1+\epsilon},
\end{equation} for algebraic convenience. The expressions in Eq. \eqref{metricAOS} reduce to those introduced in, e.g., Ref. \cite{Properties}. As discussed in that same work, the effective metric functions above reduce to the standard Schwarzschild ones when quantum corrections become negligible, i.e., as the parameters $\delta_b$ and $L_{0}\delta_c$ approach zero (limit that in fact corresponds to a vanishing area gap or, equivalently, a large black hole mass). Finally, let us add that the tortoise coordinate $r_{\star}$ is now defined by $dr_{\star}= r_s \sqrt{F_{\text{AOS}}(R)/G_{\text{AOS}}(R)}\, dR$.

For gravitational perturbations about the effective AOS model, the axial potential in the dressed metric approach takes then the expression (for further details, see Ref. \cite{CorralOlmedo})
\begin{equation}\label{draxial}
    \,^{\text{(D)}}V_l^{\text{Ax}}(R)=G_{\text{AOS}}(R)\left[\frac{l(l+1)}{H_{\text{AOS}}(R)}-R_{\text{AOS}}(R)\right], 
\end{equation}
where, suppressing the dependence on the $R$ coordinate of the AOS functions,
\begin{equation}
    R_{\text{AOS}}=\frac{2}{H_{\text{AOS}}} + \frac{1}{r_s^{2}F_{\text{AOS}}} \left[ \frac{\partial_R H_{\text{AOS}} }{H_{\text{AOS}}} \left( \frac{\partial_R G_{\text{AOS}}  }{4G_{\text{AOS}}} -\frac{\partial_R F_{\text{AOS}}}{4F_{\text{AOS}}}-\frac{3\partial_R H_{\text{AOS}}}{4H_{\text{AOS}}}\right) +\frac{\partial_R^2 H_{\text{AOS}}}{2H_{\text{AOS}}}\right].
\end{equation} 
On the other hand, the potential for the polar sector (suppressing once again the dependence on the $R$ coordinate of the AOS functions) is
\begin{equation}\label{drpolar}
     \,^{\text{(D)}}V_l^{\text{Po}}(R)= \frac{G_{\text{AOS}}\,\Lambda^2}{(\Lambda+H_{\text{AOS}}R_{\text{AOS}})^2} \!\left[\frac{\left(\Lambda+2\right)}{H_{\text{AOS}}} + R_{\text{AOS}}+\frac{H_{\text{AOS}}R^2_{\text{AOS}}}{\Lambda^2}\left( \Lambda+\frac{H_{\text{AOS}}R_{\text{AOS}}}{3}\right)\right], 
\end{equation}
where $\Lambda=(l-1)(l+2)$. 
\vspace{-1em}

\subsection{Axial and polar potentials for the effective AOS model in the hybrid approach}\label{subsec3b}
An alternative prescription that has been thoroughly explored in the LQC literature is the hybrid approach \cite{hybrid1,hybrid2,hybrid5,EliMenRev}. Formally, this proposal combines two different quantum representations: one for the homogeneous background sector (for instance, a polymeric representation) and another for the perturbative degrees of freedom, typically a more standard Fock representation. The approach is rather general and can also be applied to the effective dynamics considered here by selecting semiclassical background states sharply peaked on the AOS effective solution, assuming that the dominant quantum geometry effects affect the background, and in the considered states the most important modifications are those influencing  the peak trajectory in phase space. For a more general formulation of the hybrid quantization strategy for perturbed non-rotating black holes beyond the specific AOS effective dynamics considered here, we refer the reader to Ref. \cite{BlackPumas}, where a master equation for the gauge invariant perturbations is derived without assuming a particular effective background geometry.

In contrast to the dressed metric approach, the perturbative Hamiltonian analysis in the hybrid approach is performed directly at the level of Eq. \eqref{line}, while the evaluation of the phase space variables along effective trajectories is postponed until the final step, after the perturbative treatment has been completed. This distinction is relevant because the resulting effective phase space dynamics is, in general, not equivalent to the {\it{classical}} dynamics evaluated on effective trajectories. As an example, consider the phase space function  $\bar{\Omega}_b=b p_b/(L_0\gamma)$, which appears in the classical counterpart of the Hamiltonian \eqref{hamconstraint}, used for the background in the derivation of Eq. \eqref{QNMs}. Classically, $\bar{\Omega}_b$ generates infinitesimal dilatations in the $b$-sector of the background phase space through Poisson brackets. However, its polymerized counterpart $\Omega_b$ ---see Eq. \eqref{Omegas}--- no longer generates exact infinitesimal dilatations, since the polymerization introduces a non-linear dependence on the phase space variables, modifying the corresponding Hamiltonian flow. 

Owing to these key differences, the potentials appearing in the master equations that govern the dynamics of axial and polar perturbations about effective AOS solutions differ from those presented in the previous subsection, namely Eqs. \eqref{draxial} and \eqref{drpolar}. On the one hand, the axial potential in the hybrid approach becomes
\begin{equation}\label{hyaxial}
\,^{\text{(H)}}V_l^{\text{Ax}}(R)= \frac{1}{p_c^2} \left[ l(l+1) \frac{p_b^2}{L_{0}^2}-6\Omega_b\Omega_c\right],
\end{equation}
where every phase space function must be evaluated along the trajectories of Eq. \eqref{eom}.\footnote{The difference in appearance with respect to Eq. \eqref{draxial} is partly due to this evaluation.} We recall that $\Omega_b$ and $\Omega_c$ were defined in Eq. \eqref{Omegas}. On the other hand, the polar potential is now given by
\begin{equation}\label{hypolar}
\,^{\text{(H)}}V_l^{\text{Po}}(R)= \frac{p_b^4}{p_c^2\,(\Lambda p_b^2+6\Omega_b\Omega_cL_{0}^2)^2} \left\{l(l+1)\Lambda^2\frac{p_b^2 }{L_{0}^2}+ \frac{3}{2}\Omega_b\Omega_c  \left[\Lambda^2+ 3 \left(\Lambda+ 4 \Omega_b\Omega_c  \frac{L_{0}^2}{p_b^2} \right)^2\right]\right\},
\end{equation}
where all phase space functions must likewise be evaluated along effective trajectories.

It is straightforward to verify that, in the classical limit where the polymerization parameters vanish, the potentials introduced above reduce to the standard Regge--Wheeler and Zerilli potentials of GR, that we sometimes refer to as the Schwarzschild potentials.\footnote{For the dressed metric prescription, the result follows directly from continuity arguments about the metric functions, given that the potentials have the same dependence on them as in GR.} Indeed, according to the expressions of $\Omega_c$ and $\Omega_b$ ---see Eq. \eqref{Omegas}--- and the definitions in Eq. \eqref{convenience}, we find that, in the classical limit where $\delta_b,L_{0} \delta_c\to 0$ (and, hence, $\epsilon \to 0$, $\chi \to R$, and $\mathscr{S}\to 2$): 
\begin{equation}
\begin{aligned}
    \frac{p_b^2}{L_{0}^2} &= \frac{r_s^2 (\chi-1)(\mathscr{S}^2\chi-\epsilon^2)(\mathscr{S}\chi+\epsilon)^2 }{16(1+\epsilon)^4\chi^2} \,\xrightarrow{\;\epsilon\to 0\;}\, r^2\left(1-\frac{2M}{r}\right),\\
    \Omega_b^2 &= \frac{r_s (\chi -1)(\mathscr{S}^2\chi-\epsilon^2)}{4(1+\epsilon)^2\chi}\,\xrightarrow{\;\epsilon\to 0\;}\, r\left( 1 -\frac{2M}{r}\right),\\
    p_c &= r^2 \left( 1+ \frac{\gamma^2 ( {\delta}_c L_{0})^2 \,r_s^2}{16r^4}\right) \,\xrightarrow{\;L_{0}\delta_c\to 0\;}\, r^2.
\end{aligned}
\end{equation}
Therefore, as we have claimed, the hybrid potentials converge to the Regge--Wheeler and Zerilli potentials, respectively given by
\begin{equation}\label{GRaxial}
    \,^{\text{(Sch)}}{V^{\text{Ax}}_l(r)} = \left(1 - \frac{2M}{r}\right) \left[ \frac{l(l + 1)}{r^2} - \frac{6M}{r^3} \right],
\end{equation}
and 
\begin{equation}\label{GRpolar}
    \,^{\text{(Sch)}}{V^{\text{Po}}_l(r)}  = \frac{r^2\left(1 - \frac{2M}{r}\right)}{(\Lambda r+6 M)^2} \left\{ \frac{\Lambda^2}{r^2} \left[ l(l + 1) + \frac{6M}{r} \right] + \frac{36M^2}{r^4} \left( \Lambda + \frac{2M}{r} \right) \right\}.
\end{equation}

\begin{figure}[!hb]
    \centering
    \includegraphics[scale=0.05]{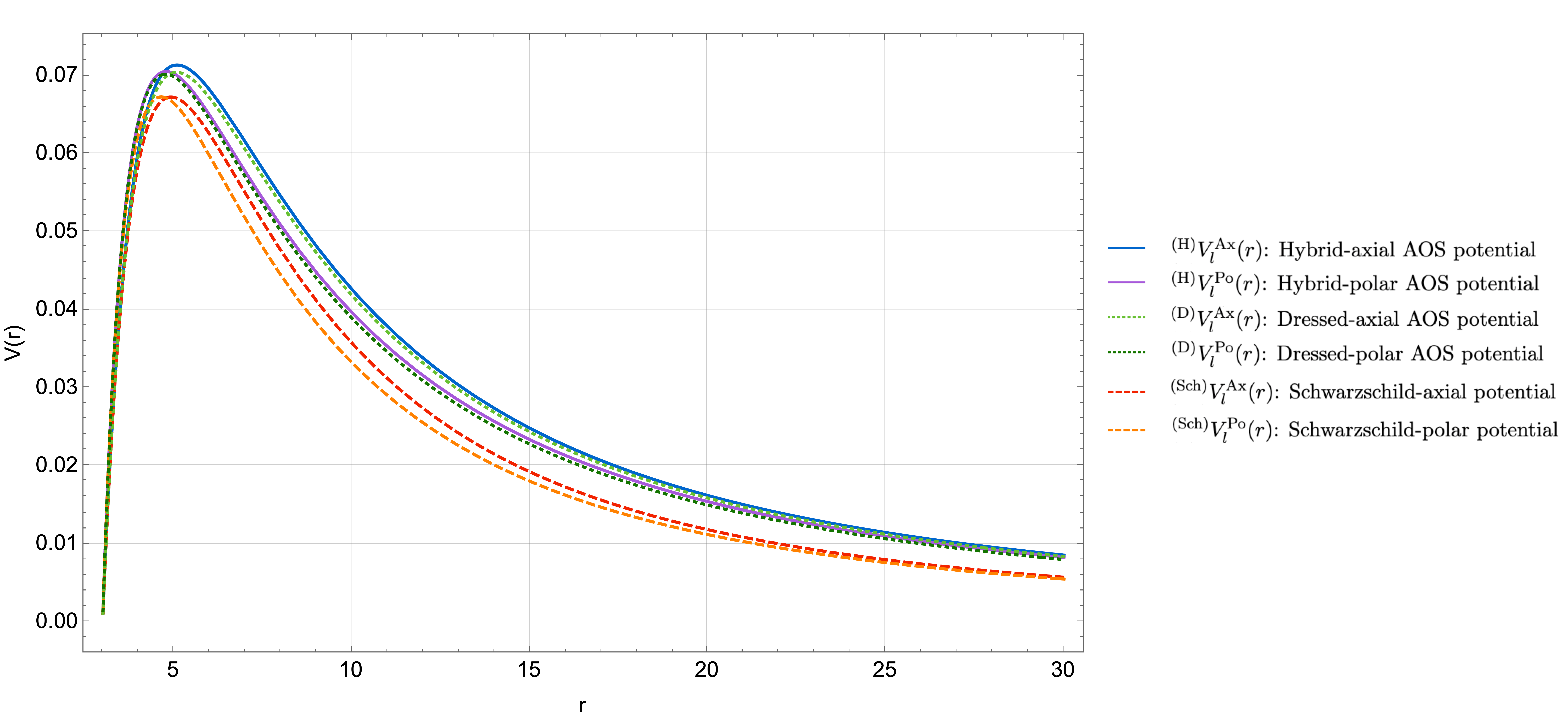}
    \caption{Comparison of the six considered potentials for $l=2$ and a black hole mass $M = 1.5$. The plot contrasts the four quantum corrected potentials for the effective AOS model against the two Schwarzschild potentials, highlighting the differences between the axial and polar sectors.}
    \label{Potentials}
\end{figure}

Before proceeding with our analysis and comparison of the quasinormal modes in the dressed metric and hybrid approaches, it seems worthwhile to take a moment to summarize the situation. Using two different prescriptions to incorporate quantum effects on the gauge invariant perturbations owing to the loop quantization of the non-rotating black hole geometry, we have obtained two distinct sets of axial and polar potentials: those derived within the dressed metric scheme ---Eqs. \eqref{draxial} and \eqref{drpolar}--- and those obtained using the hybrid approach ---Eqs. \eqref{hyaxial} and \eqref{hypolar}. In addition to these four potentials for the effective AOS model, we also consider their common classical limits for the Schwarzschild solution, namely the standard Regge--Wheeler and Zerilli potentials given by Eqs. \eqref{GRaxial} and \eqref{GRpolar}, respectively. Therefore, our analysis involves a total of six different potentials. To illustrate their differences, we have plotted these potentials in Fig. \ref{Potentials}. Although the reference mass used in the computation of quasinormal modes in the following sections is $M=500$ in Planck units (motivated by the choice made in Ref. \cite{CorralOlmedo}), for this plot of the potentials we instead employ a mass of $M=1.5$ in order to make the differences between them more apparent. For the same reason, we plot the potentials for $l=2$, since we have noticed that for large $l$ the potentials of the effective AOS model rapidly approach their shared classical counterparts. Finally, let us point out that the Schwarzschild potentials are the ones in which both the axial and polar peaks seem to be placed at the same height. 
\vspace{-1em}

\section{Semianalytical WKB method and quasinormal spectrum\label{sec4}}
In this section, we briefly discuss the WKB procedure for the computation of quasinormal modes. For more details on the method, see Refs. \cite{KZZ, ShutzWill, IyerWill, Opala, CorralOlmedo}.
One of the primary motivations for using this method is due to the similarities between the mode equations for the black hole perturbations and the one-dimensional Schrödinger equation with a potential barrier, where the WKB approximation is a powerful tool to compute reflected and transmitted amplitudes. To address this kind of problem in gravitational physics, the method was first proposed by Schutz and Will \cite{ShutzWill}, pointing out an important distinction when studying quasinormal modes, namely the boundary conditions discussed in Sec. \ref{sec3}. These lead us to consider only selected scenarios where the WKB approximation is applicable, that is, when there is a small region between the turning points of the potential, namely the points where $\left[\tilde{\omega}_{n,l}-V_{l}(r)\right]=0$. To address broader scenarios,  the modification implemented by Schutz and Will, and later generalized to higher orders by Iyer and Will \cite{IyerWill}, and by Matyjasek and Opala \cite{Opala}, is to separate the potential into three regions. The first and third regions correspond to the regions near the horizon and close to spatial infinity, respectively, where asymptotic WKB solutions can be extracted. Then, these solutions are imposed to match the solution obtained in the second and intermediate region, containing the turning points, where the potential is Taylor approximated at its peak to a given order.

After imposing the boundary conditions on the matched 
solution, we obtain an asymptotic expression for quasinormal modes. Following Refs. \cite{Opala, CorralOlmedo}, this gives
\begin{equation}\label{asymptotic}
    \tilde{\omega}_{n,l}^2 = V_{l}(r_{_\barwedge})-\sqrt{-2 \frac{d^2V_{l}(r_{_\barwedge})}{dr^2}} \left[\left(n+\frac{1}{2}\right) + \sum_{j=2}^{N} \mathcal{A}_{n,l}^{(j)}(r_{_\barwedge}) \right],
\end{equation}
where $r_{_\barwedge}$ is the peak of the potential $V_{l}$ and $\mathcal{A}_{n,l}^{(j)}(r_{_\barwedge})$ contains higher-order WKB corrections that purely depend on the overtone $n$, the multipole number $l$, and the derivatives of the potential at its peak. Further details about these functions can be found, e.g., in Refs. \cite{IyerWill, Opala}. Although the WKB mechanism can provide a semianalytical result to the problem, it nevertheless brings along an additional issue. The fact that Eq. \eqref{asymptotic} is an asymptotic expansion already suggests that the series may not converge. The role of the method is precisely to extend the WKB asymptotic solutions across the turning points. This extension may not be analytical at those points. In this sense, the turning points can be viewed as singularities where the method itself breaks down. It is therefore natural to expect that the resulting asymptotic series (which encodes information about a solution throughout all three regions) in general does not converge, a fact that is later confirmed numerically. The remaining open question is then which WKB order should be regarded as providing an optimal approximation, a choice that depends on the specific case, particularly on the curvature and asymptotic behavior of the potentials.

For this reason, using the code provided in Ref. \cite{numericalcode}, we have computed the quasinormal modes up to the fifteenth order for each potential. Previous literature and our own numerical tests confirm that the method performance improves drastically when $l\gg n$. We restrict our analysis to $n=0$ and $l=2$, which should sit on the threshold of numerical stability where the method is on the verge of breaking down. Finding a well-suited WKB order for these parameters can be seen as a stress test, ensuring the method validity for higher angular momentum numbers $l$ and different overtones $n$ where the numerics behave significantly better. The criterion we employ to identify the optimal WKB order is to select the highest one before the complex value of the quasinormal mode starts to show a significant change or oscillate. 

\begin{figure}[!h]
    \centering
        \centering
        \includegraphics[scale=0.08]{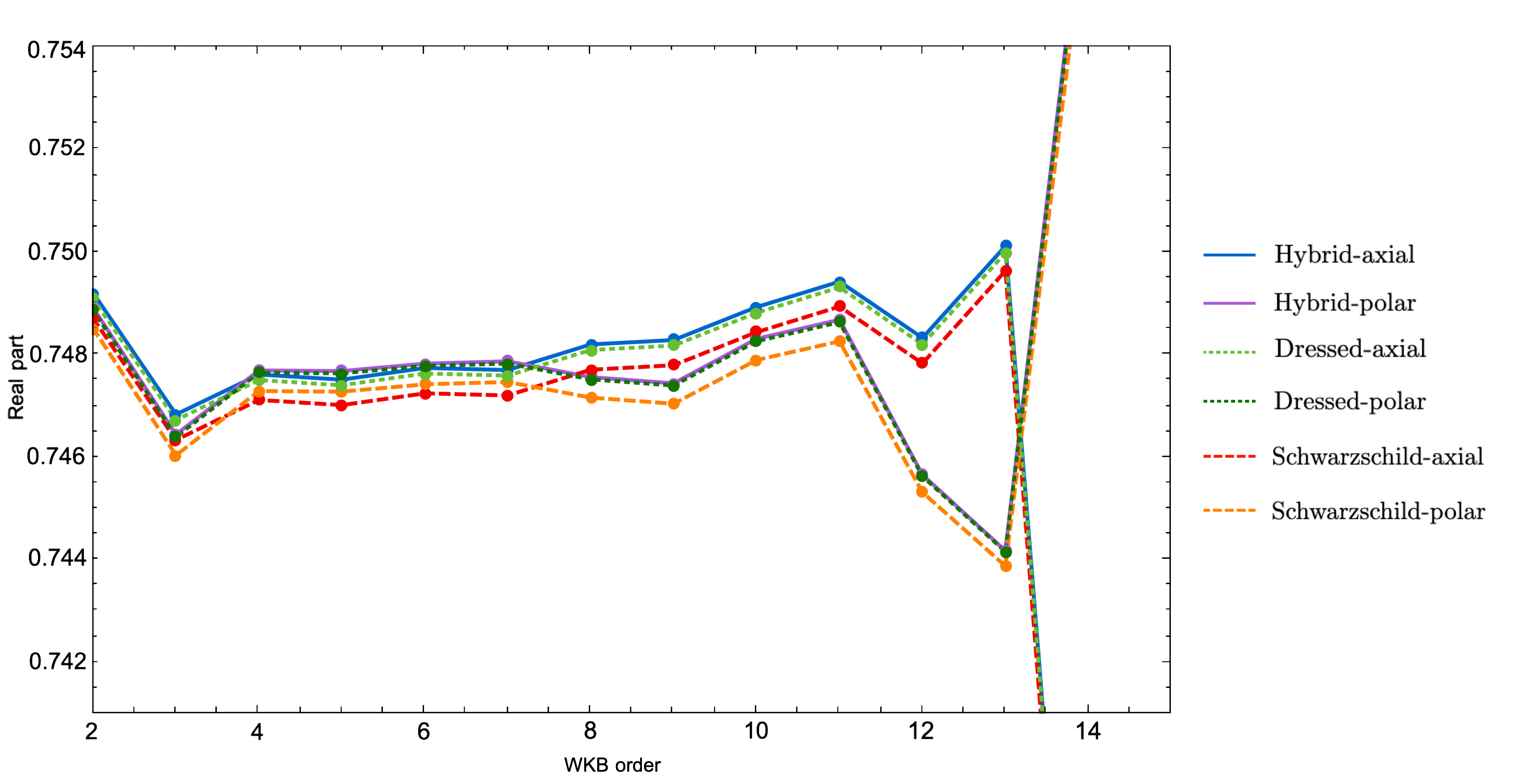}
        \caption{Real part of the quasinormal mode frequency $\omega_{0,2}$ computed using WKB orders from second to fifteenth. This plot highlights the optimal WKB order by showing where the frequency value stays stable before breakdown occurs.}
        \label{RealFreqWKB}
\end{figure}

\begin{figure}[!h]
    \centering
        \includegraphics[scale=0.08]{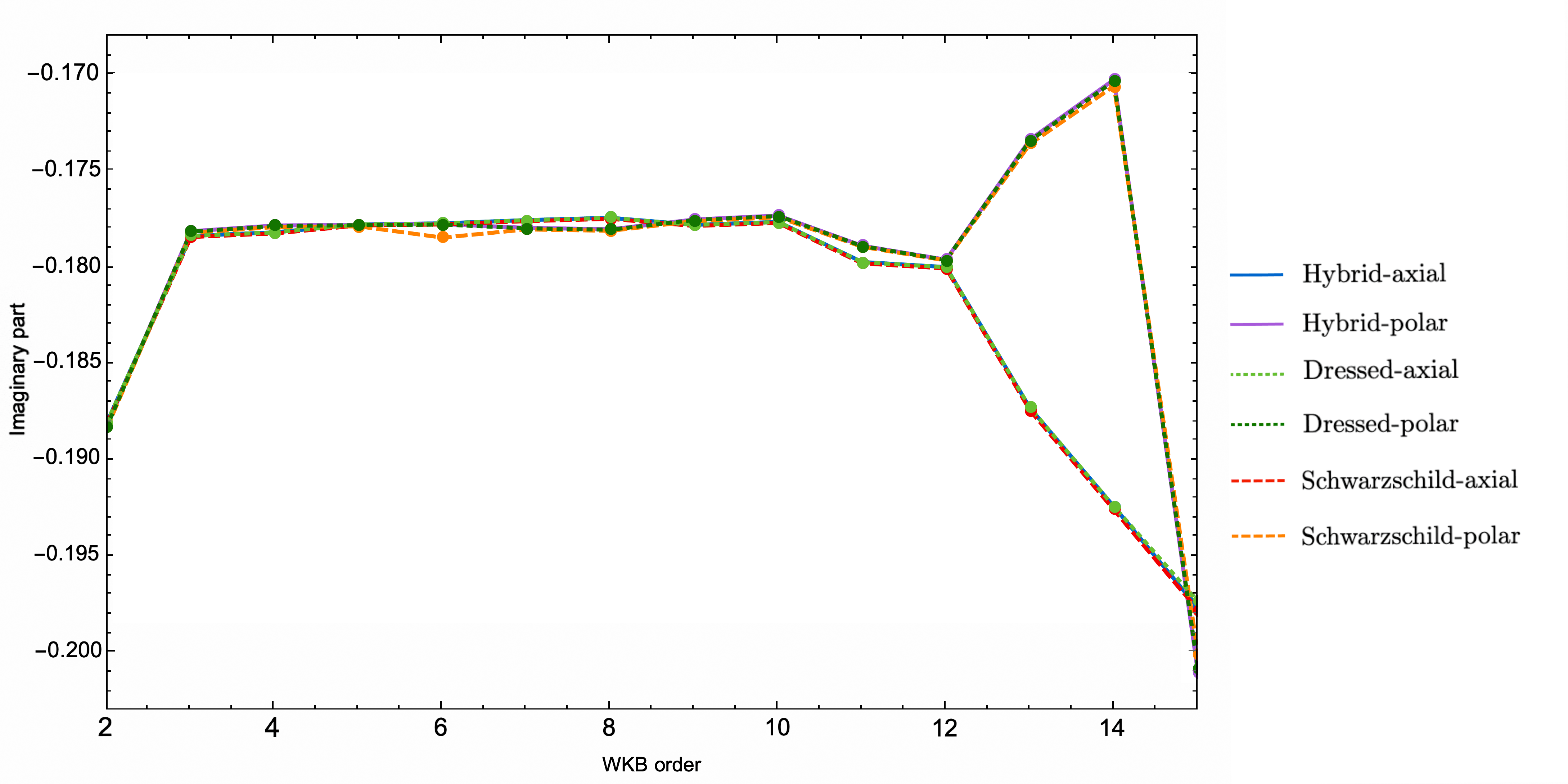}
        \caption{Imaginary part of the quasinormal mode frequency $\omega_{0,2}$ computed using WKB orders from second to fifteenth. This plot shows which WKB order gives an optimal frequency value before the method becomes unstable.}
        \label{ImaFreqWKB}
\end{figure}

Figures \ref{RealFreqWKB} and \ref{ImaFreqWKB} respectively show the real and imaginary parts for both axial and polar quasinormal modes with $n=0$ and $l=2$ in all studied cases. For the Schwarzschild and dressed metric cases, we selected a tenth WKB order in the approximation. While the data support this choice, it was primarily motivated by the desire to maintain consistency with the results presented in Ref. \cite{CorralOlmedo}. Similarly, data for this critical case indicate that the tenth WKB order is best suited for the hybrid polar potential. Conversely, for the axial hybrid case, we opted for an eleventh WKB order. Had we selected the tenth order instead, the isospectrality breaking observed in our results would have become even more prominent (after the resummation refinement discussed in Sec. \ref{subsec4a}). Since our goal is to qualitatively determine the presence of isospectrality breaking rather than to quantify its exact magnitude, we adopted the more conservative eleventh-order approach. Crucially, if violation of isospectrality is established under this conservative choice, its presence is guaranteed in the more prominent tenth-order case as well. In this manner, we safely take into account the inherent semianalytical and numerical errors when computing these modified quasinormal modes. Table \ref{table1} lists the orders chosen, based on the aforementioned criterion for the six different potentials, accompanied by the resulting values for the respective quasinormal mode frequencies. In all our tables, the reported values for the quasinormal mode frequencies $\omega_{n,l}$ are dimensionless, defined using the inverse of the Schwarzschild radius, $\omega_{n,l}=r_s\,\tilde{\omega}_{n,l}$.
\vspace{-1em}
\begin{table}[H]
\centering
\caption{Choices of WKB order for the potentials and their corresponding quasinormal mode frequency for $n=0$ and $l=2$.}
\label{table1}
\begin{tabular}{|l|c|l|}
\hline
\multicolumn{1}{|c|}{\textbf{Potential}} & \textbf{WKB order} & \multicolumn{1}{c|}{\textbf{$\mathbf{\omega}_{0,2}$}} \\ \hline
Hybrid-axial        & 11 & $0.74942073 - 0.17973356\,i$ \\ \hline
Hybrid-polar          & 10 & $0.74830458 - 0.17732003\,i$ \\ \hline
Dressed-axial       & 10 & $0.74881229 - 0.17765727\,i$ \\ \hline
Dressed-polar        & 10 & $0.74825738 - 0.17734657\,i$ \\ \hline
Schwarzschild-axial  & 10 & $0.74789066 - 0.17741012\,i$ \\ \hline
Schwarzschild-polar  & 10 & $0.74789066 - 0.17741012\,i$ \\ \hline
\end{tabular}
\end{table}

\subsection{Refinement through Padé approximants\label{subsec4a}}

Within the context of gravitational waves, it is well known that a greater increase in accuracy for quasinormal mode frequencies is achieved by introducing Padé approximants. The algorithm incorporated in the Mathematica code works as follows. For a given WKB order $u$ (say the tenth order, as an example used in our computations), the code first computes all possible Padé approximants $P_{\alpha/\beta}$ with integers $\alpha$ and $\beta$ satisfying $\alpha+\beta=u$, yielding $u+1$ distinct approximants. Since two near-identical approximants typically point to a reliable, converged result, the algorithm compares nearby pairs by calculating their relative mean values,
\begin{equation}
\bar{\omega}_{\alpha/\beta}^{(p)}=\frac{\tilde{\omega}_{(\alpha+p)/(\beta-p)}+\tilde{\omega}_{\alpha/\beta}}{2},\end{equation}
where $p$ is an integer number that determines the proximity of the pair. Because approximants near the diagonal (i.e., $\alpha \approx \beta$) are typically the most stable, the code for this case evaluates pairs up to $p=4$. For each $p$, the algorithm then chooses the two estimates $\bar{\omega}_{\alpha/\beta}^{(p)}$ for which the relative difference of their corresponding Padé approximants is minimum. It is worth noting that, under this selection criterion, the algorithm can choose the same stable Padé approximant more than once during the pairwise comparisons. In total, for the case with $p=4$ under
consideration, the algorithm selects eight values (four pairs). The final quasinormal mode frequency is then computed as their average, which effectively becomes a weighted average if any single approximant has been selected multiple times.

\begin{table}[H]
\centering
\caption{Dimensionless Schwarzschild quasinormal mode frequencies for a reference mass $M=500$ in Planck units. The third column lists the complex norm of the difference between the axial and polar frequencies, while the fourth column gives the square root of the sum of squares of the individual axial and polar error estimates.}
\label{GRtable}
\begin{tabular}{|l|c|c|c|c|}
\hline
\multicolumn{5}{|c|}{Schwarzschild quasinormal mode frequencies $\omega_{n,l}$} \\ \hline
(n, l) & Axial & Polar & Difference & Error \\ \hline
(0,2) & $0.74733225-0.17792806\,i$ & $0.74734291-0.17792545\,i$ & $1.0975 \times 10^{-5}$ & $6.3922 \times 10^{-6}$ \\ \hline
(1,2) & $0.69322645-0.54811740\,i$ & $0.69337238-0.54785794\,i$ & $2.9768 \times 10^{-4}$ & $4.7479 \times 10^{-4}$ \\ \hline
(0,3) & $1.19888658-0.18540612\,i$ & $1.19888656-0.18540609\,i$ & $3.6056 \times 10^{-8}$ & $4.5884 \times 10^{-8}$ \\ \hline
(1,3) & $1.16528891-0.56259764\,i$ & $1.16528740-0.56259608\,i$ & $2.1711 \times 10^{-6}$ & $3.1567 \times 10^{-7}$ \\ \hline
(0,4) & $1.61835676-0.18832792\,i$ & $1.61835675-0.18832792\,i$ & $1.0000 \times 10^{-8}$ & $6.7341 \times 10^{-10}$ \\ \hline
(1,4) & $1.59326305-0.56866870\,i$ & $1.59326306-0.56866870\,i$ & $1.0000 \times 10^{-8}$ & $1.8007 \times 10^{-8}$ \\ \hline
\end{tabular}
\end{table}

Table \ref{GRtable} presents the corresponding axial and polar quasinormal modes for a black hole with mass $M = 500$. Since the spectrum of the Schwarzschild quasinormal modes is theoretically known to be isospectral, comparing them provides a notion of the reliability and accuracy of the employed WKB method, though this accuracy naturally varies from one potential to another. While the numerical error estimation provided by the Mathematica code is not perfectly precise, it serves as an indicative guide. To quantify these uncertainties, the values in the third column are computed as the complex norm of the difference between the axial and polar quasinormal modes, while the last column is the geometric sum (the square root of the sum of squares) of the individual axial and polar error estimates according to Mathematica.

\begin{table}[ht!]
\small
\centering
\caption{Dimensionless dressed quasinormal mode frequencies for a reference mass $M=500$ in Planck units. The difference in the third column is computed as in Table \ref{GRtable}. The fourth column gives the error as the square root of the sum of the squared axial and polar error estimates and of the squared Schwarzschild difference. The last column gives the ratio of the difference to the error.}
\label{dressed_qnm}
\begin{tabular}{|l|c|c|c|c|c|}
\hline
\multicolumn{6}{|c|}{Dressed metric quasinormal mode frequencies $\omega_{n,l}$} \\ \hline
(n, l) & Axial & Polar & Difference & Error & Isospectrality \\ \hline
(0,2) & $0.74771876-0.17787563\,i$ & $0.74770137-0.17786831\,i$ & $1.8868 \times 10^{-5}$ & $1.2689 \times 10^{-5}$ & $1.4869$ \\ \hline
(1,2) & $0.69360839-0.54794196\,i$ & $0.69371455-0.54766125\,i$ & $3.0011 \times 10^{-4}$ & $5.5841 \times 10^{-4}$ & $0.5374$ \\ \hline
(0,3) & $1.19942465-0.18535672\,i$ & $1.19941685-0.18535615\,i$ & $7.8208 \times 10^{-6}$ & $5.7149 \times 10^{-8}$ & $136.85$ \\ \hline
(1,3) & $1.16581711-0.56244359\,i$ & $1.16580600-0.56243957\,i$ & $1.1815 \times 10^{-5}$ & $2.1946 \times 10^{-6}$ & $5.3837$ \\ \hline
(0,4) & $1.61905472-0.18828092\,i$ & $1.61905146-0.18828080\,i$ & $3.2622 \times 10^{-6}$ & $1.0022 \times 10^{-8}$ & $325.50$ \\ \hline
(1,4) & $1.59395164-0.56852466\,i$ & $1.59394788-0.56852416\,i$ & $3.7931 \times 10^{-6}$ & $2.0015 \times 10^{-8}$ & $189.51$ \\ \hline
\end{tabular}
\end{table}

Tables \ref{dressed_qnm} and \ref{hybrid_qnm} present the corresponding dimensionless axial and polar quasinormal mode frequencies for the dressed metric and hybrid prescriptions, respectively, using the same black hole reference mass, $M = 500$. The differences shown in these tables are computed using the same method as before. However, to account for the intrinsic systematic error of the WKB method inherited from the Schwarzschild scenario, the error in the fourth column is calculated as the square root of the sum of the squared axial error estimate, the squared polar error estimate, and the squared complex norm of the Schwarzschild difference. 

\begin{table}[ht!]
\small
\centering
\caption{Dimensionless hybrid quasinormal mode frequencies for a reference mass $M=500$ in Planck units. The difference in the third column is computed as in Table \ref{GRtable}, whereas the error in the fourth column and the isospectrality ratio in the last column are calculated as in Table \ref{dressed_qnm}.}
\label{hybrid_qnm}
\begin{tabular}{|l|c|c|c|c|c|}
\hline
\multicolumn{6}{|c|}{Hybrid quasinormal mode frequencies $\omega_{n,l}$} \\ \hline
(n, l) & Axial & Polar & Difference & Error & Isospectrality \\ \hline
(0,2) & $0.74779134-0.17786326\,i$ & $0.74774536-0.17784449\,i$ & $4.9664 \times 10^{-5}$ & $1.4947 \times 10^{-5}$ & $3.3228$ \\ \hline
(1,2) & $0.69366092-0.54769395\,i$ & $0.69375147-0.54758180\,i$ & $1.4414 \times 10^{-4}$ & $3.1530 \times 10^{-4}$ & $0.4571$ \\ \hline
(0,3) & $1.19948799-0.18535267\,i$ & $1.19946022-0.18534799\,i$ & $2.8161 \times 10^{-5}$ & $4.6210 \times 10^{-8}$ & $609.41$ \\ \hline
(1,3) & $1.16587987-0.56242880\,i$ & $1.16584963-0.56241385\,i$ & $3.3734 \times 10^{-5}$ & $2.5625 \times 10^{-6}$ & $13.164$ \\ \hline
(0,4) & $1.61910026-0.18827818\,i$ & $1.61908819-0.18827669\,i$ & $1.2161 \times 10^{-5}$ & $1.0002 \times 10^{-8}$ & $1215.8$ \\ \hline
(1,4) & $1.59399769-0.56851610\,i$ & $1.59398490-0.56851144\,i$ & $1.3612 \times 10^{-5}$ & $1.3356 \times 10^{-8}$ & $1019.1$ \\ \hline
\end{tabular}
\end{table}
\begin{figure}[H]
    \centering
    \includegraphics[scale=0.085]{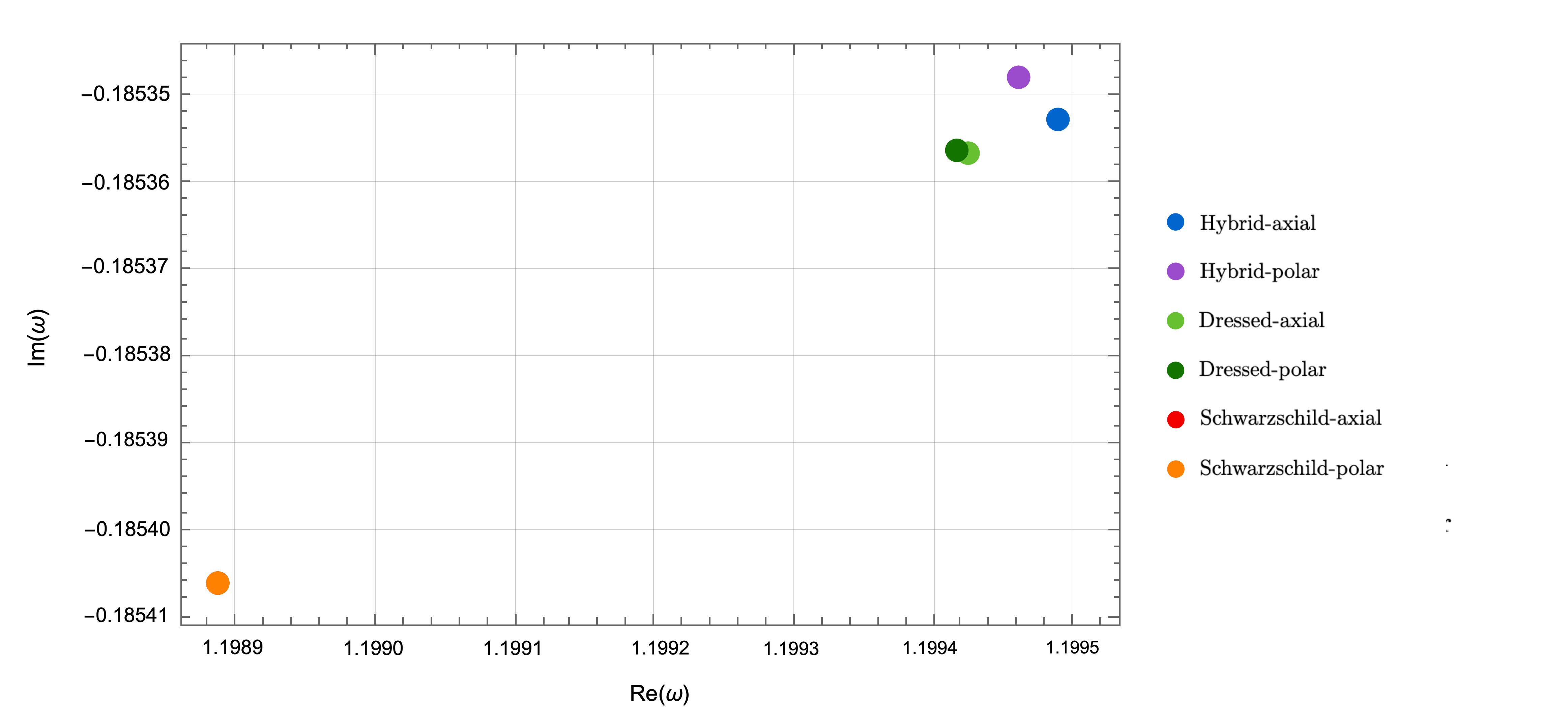}
    \caption{Quasinormal mode frequencies for the mode with $n=0$ and $l=3$ of the Schwarzschild, dressed metric, and hybrid potentials. While the Schwarzschild case exhibits perfect isospectrality at this scale, a clear violation of isospectrality is observed for the two cases with quantum corrections.}
    \label{iso}
\end{figure}

In addition, to measure the difference between the axial and polar sectors and elucidate whether there is a genuine violation of isospectrality or the observed differences are merely due to the numerical uncertainty of the method, in Tables \ref{dressed_qnm} and \ref{hybrid_qnm} we append an \textit{Isospectrality} column, defined as the ratio of the difference (in complex norm) to the combined error estimate. A ratio between 0 and 1 for most values would suggest that isospectrality is preserved within numerical precision; a value between 1 and 5 is likely to be inconclusive; and a value above 5 strongly indicates a physical breaking of isospectrality. For almost all evaluated modes, this ratio lies well above 5. Remarkably, approaching the eikonal limit (i.e., $l\gg n$) where the WKB accuracy is highest, every isospectrality ratio points towards violation of isospectrality. Conversely, for the $n=1$ and $l=2$ mode, the ratio falls below 1, superficially suggesting that isospectrality holds. As will be clarified in the next section, however, this specific case actually marks a complete breakdown of the WKB method itself, rather than a preservation of the symmetry.

\begin{figure}[H]
    \centering
    \includegraphics[scale=0.032]{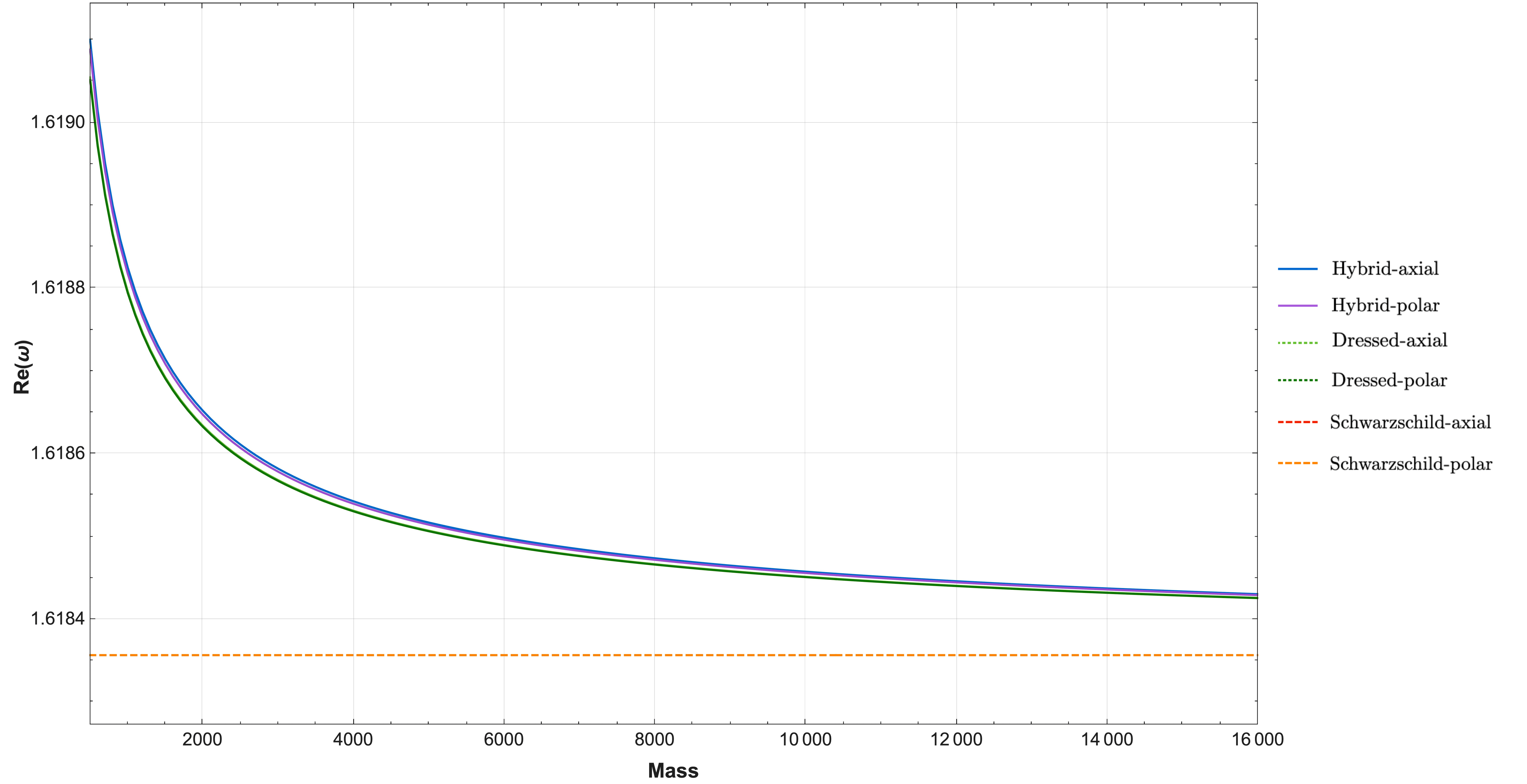}
    \caption{Real part of the dimensionless quasinormal mode frequency $\omega_{0,4}$ for the dressed metric and hybrid prescriptions as a function of the black hole mass (in Planck units). As the mass increases, the real parts in both prescriptions asymptotically converge to the standard GR value, which remains constant owing to the scale invariance of the classical Schwarzschild potentials.}
    \label{realMass}
\end{figure}
\begin{figure}[H]
    \centering
     \includegraphics[scale=0.049]{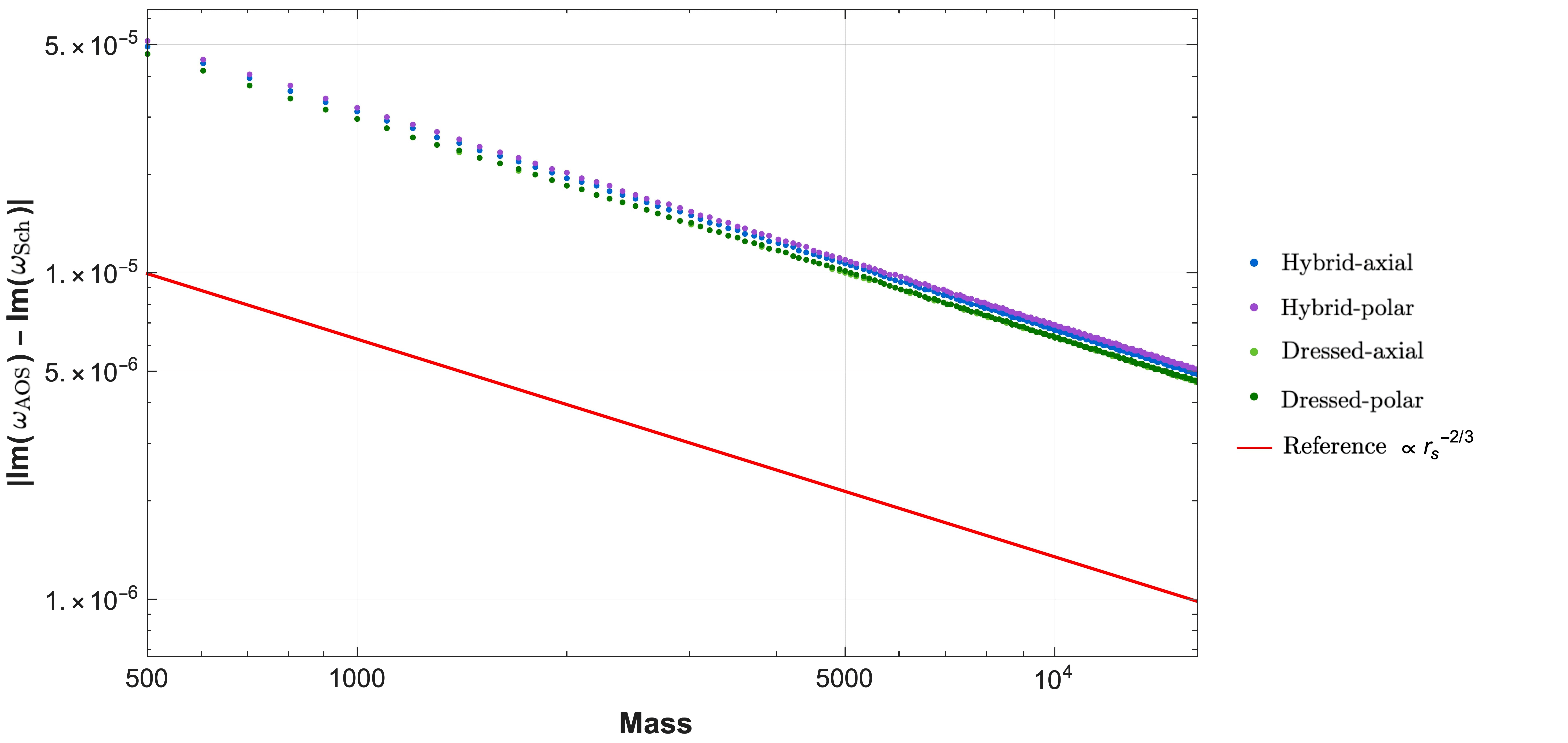}
        \caption{Difference in norm of the imaginary parts of the quasinormal mode frequencies for $n=0$ and $l=4$ with respect to that of the Schwarzschild solution, $\omega_{\text{Sch}}$, plotted against the black hole mass using a log-log scale. The plot displays this difference for both the axial and polar sectors and for both the hybrid and dressed metric cases. The real parts exhibit an entirely analogous behavior. A reference line equal to $10^{-3} r_s^{-2/3}$ is included for comparison.}
        \label{LogLogGrDev}
\end{figure}

In Fig. \ref{iso}, where the overtone and angular momentum numbers are $n=0$ and $l=3$, respectively, the graphics clearly illustrate the violation of isospectrality for both the dressed metric and hybrid prescriptions in the effective AOS model. In contrast, isospectrality is clearly preserved at that scale for the standard Schwarzschild case. On the other hand, Fig. \ref{realMass} shows the behavior of the real parts of the dimensionless quasinormal mode frequencies with $n=0$ and $l=4$ for both the dressed metric and hybrid prescriptions as the black hole mass varies. As expected, the values for both prescriptions asymptotically converge to the GR value at large masses. This GR reference value remains completely constant across the mass range, owing to the scale invariant nature of the classical Schwarzschild potentials. Moreover, we confirm the observation of Ref. \cite{CorralOlmedo} that deviations from GR generally decrease as $r_s^{-2/3}$ for both the real and imaginary parts of the quasinormal mode frequencies (see Fig. \ref{LogLogGrDev}), suggesting that these differences may be negligible for macroscopic (e.g., astrophysical) black holes. However, a more rigorous study of the topic would certainly be worthwhile.

\begin{figure}[H]
    \centering
    \includegraphics[scale=0.054]{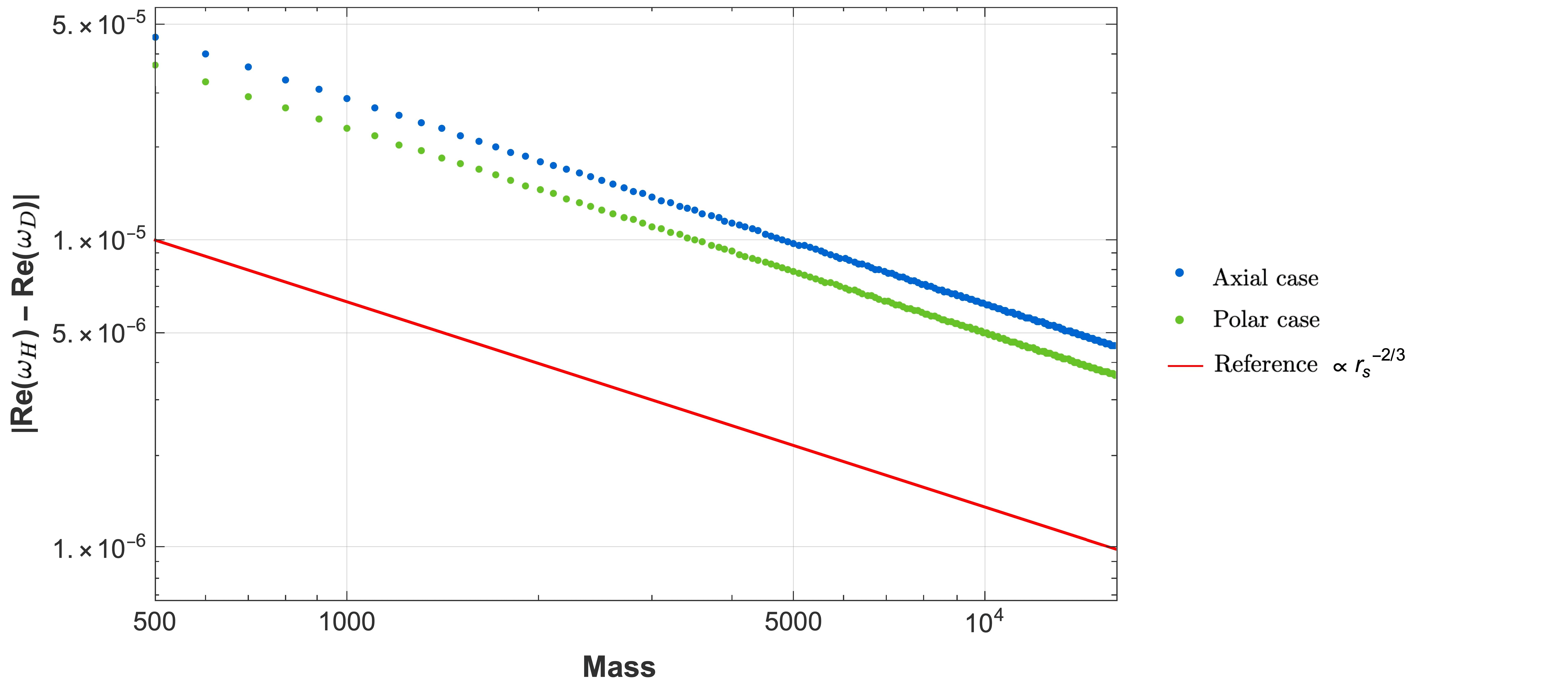}
        \caption{Difference in norm between the real parts of the dressed metric and hybrid quasinormal mode frequencies, $\omega_H$ and $\omega_D$ respectively, with $n=0$ and $l=4$,  plotted against the black hole mass on a log-log scale. The plot includes the differences for both the axial and polar sectors. The imaginary parts exhibit an entirely analogous behavior. A reference line equal to $10^{-3} r_s^{-2/3}$ is included for comparison.}
        \label{LogLogQuantDif}
\end{figure}

Finally, Fig. \ref{LogLogQuantDif} highlights the difference in a log-log plot between the real parts of the dressed and hybrid quasinormal mode frequencies with $n=0$ and $l=4$ as a function of the mass for both the axial and polar sectors. The imaginary parts can be seen to exhibit an entirely analogous behavior. 

\section{Validity range and breakdown of the WKB approximation\label{sec5}}

\begin{figure}[!h]
    \centering
    \includegraphics[scale=0.08]{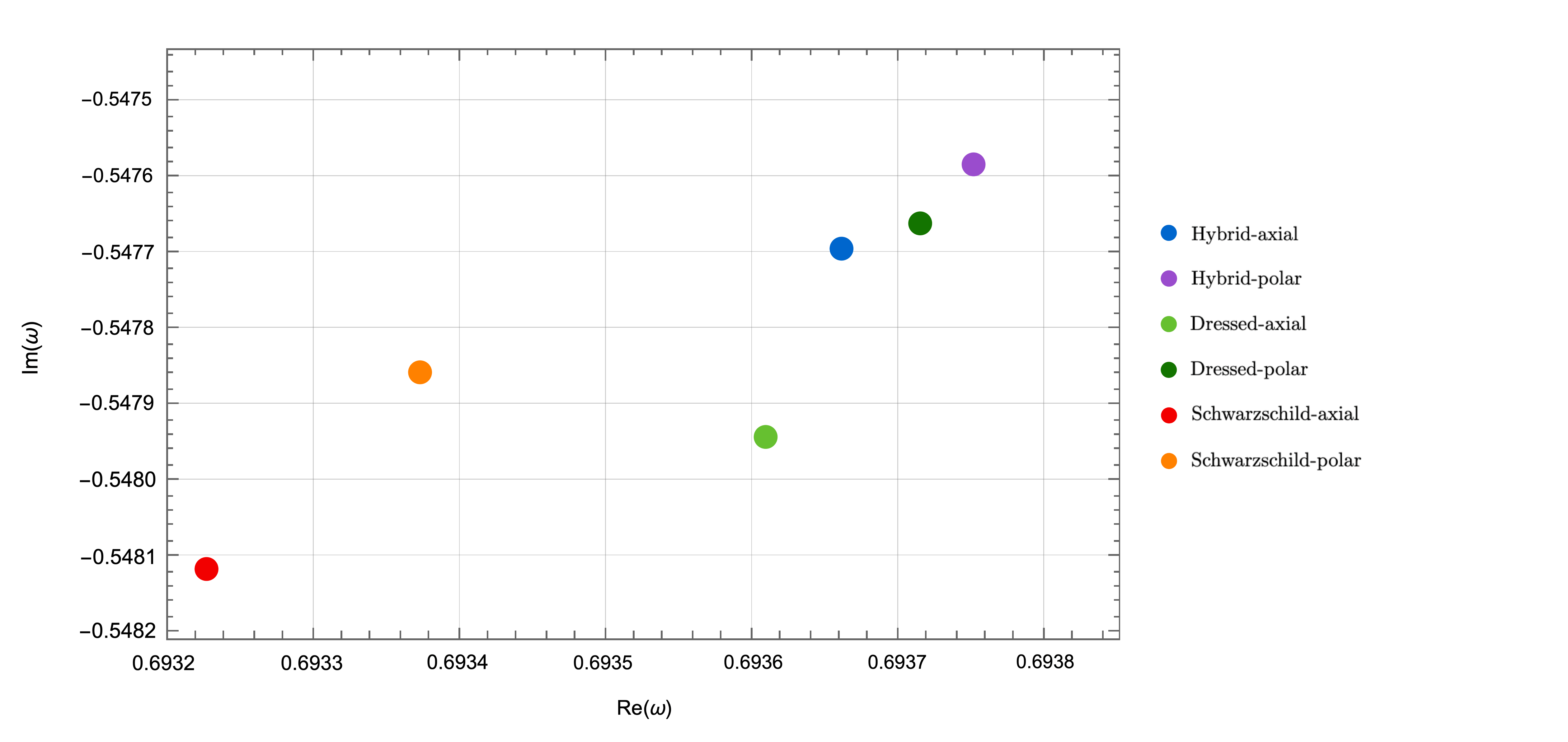}
    \caption{Quasinormal mode frequencies for the $n=1$ and $l=2$ case using a reference mass of $M=500$ in Planck units. The highly scattered distribution of points and the lack of alignment between the theoretically isospectral Schwarzschild modes highlight a breakdown of the WKB numerical method for this highly excited low-angular-momentum state.}
    \label{fig:wkb_breakdown_n1_l2}
\end{figure}

To evaluate the validity and limitations of the employed WKB method, we turn our attention to the low-angular-momentum modes, where the approximation is expected to struggle. The case $n=1$ and $l=2$ (again with a reference mass of $M=500$) provides an explicit illustration of the breakdown of the method. As shown in Fig. \ref{fig:wkb_breakdown_n1_l2}, the calculated frequencies appear as scattered points; even the standard Schwarzschild modes, which are analytically guaranteed to be isospectral, fail to cluster together. 

\begin{figure}[!h]
    \centering
    \includegraphics[scale=0.032]{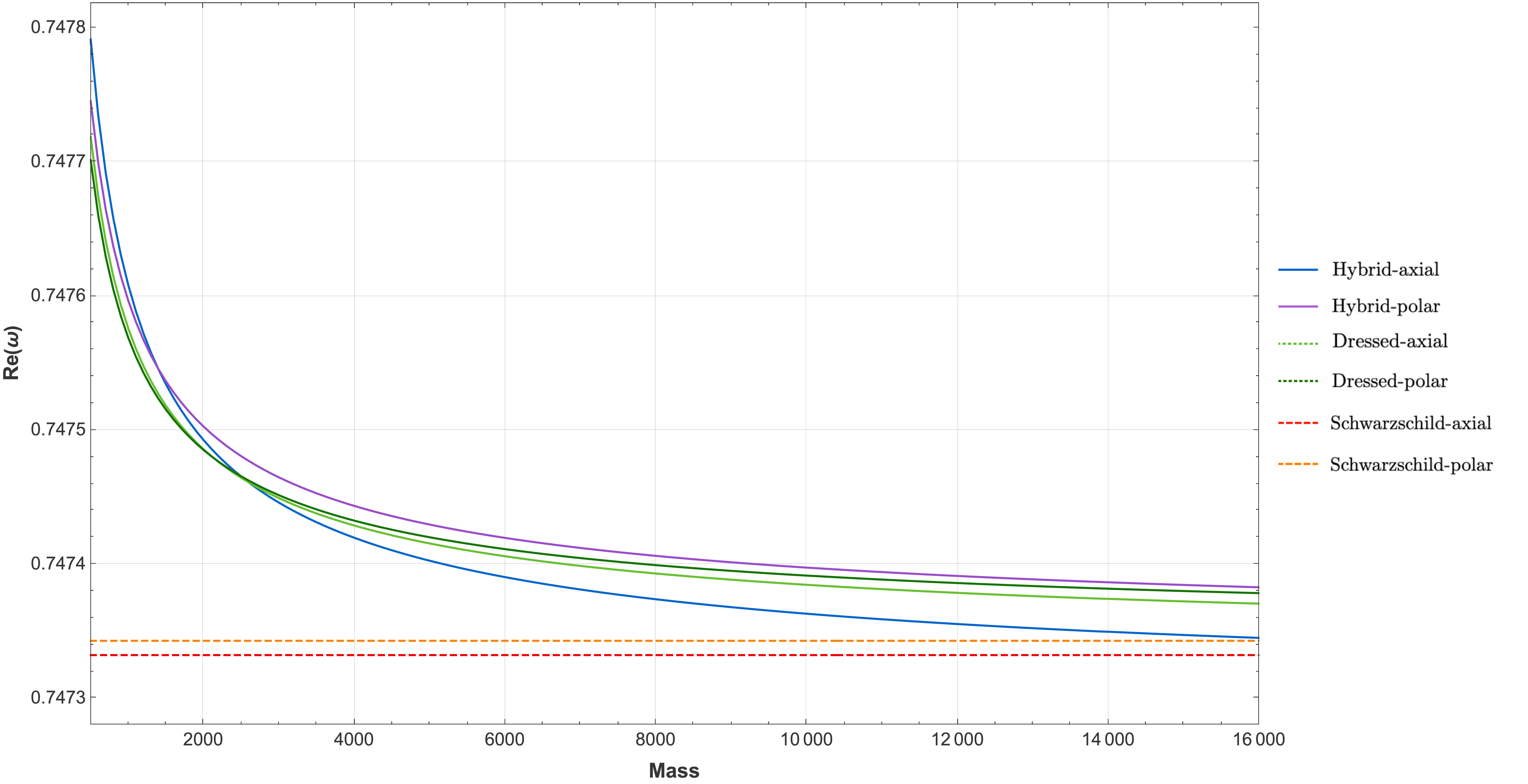}
    \caption{Mass dependence of the real parts of the quasinormal mode frequencies for $n=0$ and $l=2$. Although the Schwarzschild curves are close to each other at this scale, they do not overlap. Together with the fact that the quasinormal mode frequencies with quantum corrections for axial and polar perturbations intersect solely at a single point, this indicates that the considered mode lies at the threshold of reliability of the WKB method.}
    \label{fig:wkb_threshold_n0_l2}
\end{figure}

The $n=0$ and $ l=2$ mode shown in Fig. \ref{fig:wkb_threshold_n0_l2} appears to be on the threshold of this breakdown. While the axial and polar Schwarzschild frequencies do not diverge wildly, they noticeably fail to overlap as they should under perfect isospectrality. Because the classical potentials are scale invariant, this violation of isospectrality remains entirely independent of the black hole mass. Furthermore, for the two prescriptions with quantum corrections corresponding to the effective AOS model, the axial and polar curves actually intersect, although only at a single, isolated mass value (i.e., they are compatible with isospectrality only for a specific mass value), and they fail to asymptotically converge to the classical GR value as the mass increases. Together, these behaviors indicate that, for $n=0$ and $l=2$, the WKB method is beginning to break down, operating past its threshold of reliability.

\newpage
\section{Conclusion \label{sec6}}

Quasinormal modes constitute one of the most promising observational probes to test GR and open new possibilities for the falsification of alternative theories, including those that predict quantum gravity effects. Departures from the classical ringdown spectrum (such as the violation of the axial-polar isospectrality of black holes) could provide valuable information about effective phenomena emerging from the quantum geometry underlying these systems. In the present work, we aim to answer the question of whether the violation of isospectrality previously found in a certain class of loop quantum black holes is robust under the adoption of other prescriptions for implementing quantum corrections effectively, while also assessing the reliability of the semianalytical WKB method employed in the computation of the corresponding spectra.

More precisely, we have focused on the study of isospectrality for the AOS effective black hole model \cite{AOS,AOS2,Properties,CorralOlmedo,ENGQMM} under two distinct quantization prescriptions for the perturbations: the so-called dressed metric \cite{dressed1,dressed2,dressed3,dressed4} and hybrid \cite{hybrid1,hybrid2,hybrid5,EliMenRev} approaches. We have computed the quasinormal modes using the semianalytic WKB method alongside its Mathematica implementation introduced in Ref. \cite{numericalcode} and discussed in Refs. \cite{Opala,KZZ}, which allowed us to make a straightforward comparison with the existing literature on some of the computed dressed metric values \cite{CorralOlmedo}. Based on the study of the asymptotic description of the quasinormal mode frequencies, we selected the eleventh WKB order for the hybrid axial potential and the tenth WKB order for the remaining potentials; namely the hybrid polar, dressed axial and polar, as well as the classical Schwarzschild potentials used for comparison (i.e., the Regge--Wheeler and Zerilli potentials). These computations were then refined using Padé approximants, selecting the best suited values, averaging the final results with the same built-in function of the Mathematica code. For a reference black hole with mass $M = 500$ in Planck units, the quasinormal modes show a sensitivity to the chosen quantization prescription with differences starting at the fifth significant digit. Within the AOS scenario, our results confirm the violation of isospectrality previously reported in Ref. \cite{CorralOlmedo} for the dressed metric approach. Furthermore, we find that the hybrid approach yields a slightly larger degree of non-isospectrality, although the increase in the violation is less than one order of magnitude. For large black hole masses, the magnitude of this violation in both prescriptions decreases as $r_s^{-2/3}$. This falloff is consistent with the assumption that boundary conditions on the quasinormal modes remain unchanged when considering effective trajectories for the background solutions with quantum corrections, and with our analytical and numerical demonstration that the axial and polar potentials within both prescriptions tend to their classical counterparts (the Regge--Wheeler and Zerilli potentials) in the classical limit. Likewise, the difference (in complex norm) between the frequencies obtained with the two prescriptions decreases according to the same power law. Concerning the numerical implementation, the error estimates generated by the code (as indicated in its documentation) should be regarded as qualitative indicators rather than strict quantitative error estimates. 

It is well established that the WKB approach generates trustworthy results in the parameter space region where $n\ll l$. Beyond this context,
two specific overtone-multipole cases were identified and discussed. The first case, that we denoted as a critical case, is characterized by $n=0$ and $l=2$. The obtained output frequencies represent a critical threshold where the WKB approximation operates near its limit of validity. Graphically, when evaluating at a five-digit accuracy, isospectrality appears to break slightly for the classical Schwarzschild potentials; the same precision level at which quantum modifications manifest in the models with corrections from the effective AOS background. Because the Regge--Wheeler and Zerilli potentials are scale invariant, this apparent rupture in GR persists independently of the black hole mass. However, because the violation of isospectrality for the GR case was considered arbitrarily small (a difference of $1.0975 \times 10^{-5}$, almost extending to the sixth digit), we used it to calibrate and select the appropriate WKB orders under the rationale that orders capable of handling this borderline case will remain stable and consistent for regimes where the WKB method is robust. The second particular case that we identified is that corresponding to $n = 1$ and $ l = 2$. Although treated as reliable in other works (e.g., Ref. \cite{CorralOlmedo}), our graphical analysis shows that this case lies outside of the range of validity of the WKB method. More concretely, the calculated frequencies appear as scattered points in the complex plane at the considered accuracy and reference mass. Moreover, they exhibit artificial compatibility with isospectrality only at  a specific isolated mass value (which differs between each pair of potentials) and fail to converge to their classical GR counterparts in the large-mass limit. Hence, by the lack of physical consistency, these results demonstrate that the WKB method is pushed beyond its range of applicability for this mode.

Let us comment in these concluding remarks that, as a first approach, and partly motivated by the strategy employed for the dressed metric prescription in Ref. \cite{CorralOlmedo}, we have followed a straightforward procedure to derive the functional form of the potentials for the perturbative master equations within the framework of the hybrid quantization. Indeed, those hybrid potentials have been calculated by a naive direct evaluation of all background phase space functions appearing in them on the effective AOS trajectories. Ultimately, this work paves the way to investigate more deeply how to properly extend (if possible) the classical Chandrasekhar transformation to the effective AOS  model, such that the related axial and polar potentials remain isospectral. This particular Darboux transformation was largely considered a hidden symmetry of the system between distinct master equations and their respective potentials. Recent works \cite{GA2} have found that, within a Hamiltonian approach to black hole perturbations, the Chandrasekhar transformation can be thought of as a background-dependent canonical transformation. Determining whether this can be extended to an effective description of spacetime would have a profound impact on the conceptual understanding of isospectrality and its violation across different prescriptions in loop quantum black hole geometries. Moreover, since the Chandrasekhar transformation can be implemented in our perturbed system before its hybrid quantization, it is obviously possible to reach an isospectral quantum description. In fact, this is the route chosen to treat axial and polar perturbations in Ref. \cite{Masterfun}. Elucidating whether we can restore isospectrality at the level of the effective formulation by means of an extension of the classical transformation, and explaining how this map changes the effective counterpart of the hybrid potentials would be enlightening and allow us to place limits in using naive effective formulas in LQC. We plan to investigate these issues in future works.

Finally, let us emphasize that the numerical study of a possible violation of isospectrality for different quantization prescriptions within the framework of LQG/LQC, such as the dressed metric and hybrid approaches discussed here for a naive effective AOS limit, provides estimates of the order of magnitude and possible range of variation that this breaking can reach when quantum ambiguities found in the loop description of the perturbations come into play. These estimates can be instrumental in determining whether observations may help us fix such ambiguities or, on the contrary, whether the final predictions of our analysis are robust against their presence.

\acknowledgments
We are grateful to Beatriz Elizaga Navascués, Andrés Mínguez-Sánchez, and Javier Olmedo for insightful discussions. A.T.-C. would also like to thank Antonio Vicente-Becerril for constructive discussions on the numerics. A.T.-C. acknowledges financial support from Comunidad Autónoma de Madrid through the PIPF-2024 fellowship (Ref. PIPF-2024/TEC-34465). G.A.M.M. and A.T.-C. acknowledge support from MCIN/AEI/10.13019/501100011033 and FSE+ under the Grant No. PID2023-149018NB-C41 of the Spanish Ministry of Science and Innovation.


\begin{thebibliography}{99.}

\bibitem{Lisaone} P. Amaro-Seoane et al. (LISA Collaboration), Laser Interferometer Space Antenna, arXiv:1702.00786.


\bibitem{SET} A. Abac et al. (Einstein Telescope Collaboration), The science of the Einstein Telescope, JCAP {\bf 03}, 081 (2026).


\bibitem{Lisatwo} K.G. Arun et al. (LISA Collaboration), New horizons for fundamental physics with LISA, Living Rev. Relativity {\bf 25}, 4
(2022).


\bibitem{LISAthree} M. Piarulli, S. Marsat, E.M. Sänger, A. Buonanno, J. Steinhoff, and N. Tamanini, Parametrized test of general relativity for LISA massive black hole binary inspirals, Phys. Rev. D {\bf 112}, 124044 (2025).

\bibitem{ETtwo} A. Begnoni, W. Del Pozzo, M. Pegorin, J. Pomper, and A. Ricciardone, Tests of general relativity with Einstein Telescope, arXiv:2511.07520.

\bibitem{Tests} E. Berti et al., Testing general relativity with present and future astrophysical observations, Class. Quantum Grav. {\bf 32}, 243001 (2015).

\bibitem{Berti} E. Berti et al., Black hole spectroscopy: From theory to experiment, Class. Quantum Grav. {\bf 43}, 123001 (2026).



\bibitem{Nollert} H.-P. Nollert, Quasinormal modes: The characteristic “sound” of black holes and neutron stars, Class. Quantum Grav. {\bf 16},
R159 (1999).

%%

\bibitem{ReggeWheeler} T. Regge and J.A. Wheeler, Stability of a Schwarzschild singularity, Phys. Rev. {\bf 108}, 1063 (1957).

\bibitem{Zerilli} F.J. Zerilli, Effective potential for even parity Regge Wheeler gravitational perturbation equations, Phys. Rev. Lett. {\bf 24}, 737 (1970).

\bibitem{Chandra} S. Chandrasekhar and S.L. Detweiler, The quasi-normal modes of the Schwarzschild black hole, Proc. Roy. Soc.
Lond. A {\bf 344}, 441 (1975).


\bibitem{Yurov} A. Yurov and V. Yurov, A look at the generalized Darboux transformations for the quasinormal spectra in Schwarzschild black hole perturbation theory: Just how general should it be? Phys. Lett. A {\bf 383}, 2571 (2019).

\bibitem{Glampe} K. Glampedakis, A.D. Johnson, and D. Kennefick, The Darboux transformation in black hole perturbation theory, Phys. Rev. D {\bf 96}, 024036 (2017).


\bibitem{AP} A. Ashtekar and P. Singh, Loop quantum cosmology: A status report, Class. Quantum Grav. {\bf 28}, 213001 (2011).

\bibitem{AB} A. Ashtekar and E. Bianchi, A short review of loop quantum gravity, Rep. Prog. Phys. {\bf 84}, 042001 (2021).

\bibitem{AshLewa} A. Ashtekar and J. Lewandowski, Background independent quantum gravity: A status report, Class. Quantum Grav. {\bf 21},
R53 (2004).

\bibitem{Thiemann} T. Thiemann, Modern Canonical Quantum General Relativity (Cambridge University Press, Cambridge, England, 2007).

\bibitem{APSLetter} A. Ashtekar, T. Pawłowski, and P. Singh, Quantum nature of the big bang, Phys. Rev. Lett. {\bf 96}, 141301 (2006).

\bibitem{APS} A. Ashtekar, T. Pawłowski, and P. Singh, Quantum nature of the big bang: Improved dynamics, Phys. Rev. D. {\bf 74}, 084003
(2006).

\bibitem{MMO} M. Mart\'{\i}n-Benito, G.A. Mena Marug\'an, and J. Olmedo, Further improvements in the understanding of isotropic loop quantum cosmology, Phys. Rev. D {\bf 80}, 104015 (2009).

\bibitem{largescale} I. Agullo and N.A. Morris, Detailed analysis of the predictions of loop quantum cosmology for the primordial power spectra, Phys. Rev. D {\bf 92}, 124040 (2015).

\bibitem{OrigAnali} B. Elizaga Navascu{\'e}s and G.A. Mena Marug{\'a}n, Analytical investigation of pre-inflationary effects in the primordial power spectrum: From general relativity to hybrid loop quantum cosmology, JCAP {\bf 09}, 030 (2021).


\bibitem{Antonio} A. Guillén, K. Langer, G.A. Mena Marug\'an, N. Rodenbücher, and A. Vicente-Becerril, Parametrization of the primordial power spectrum in loop quantum cosmology, Phys. Rev. D {\bf 113}, 126026 (2026).


%% Perturbations AOS dressed


\bibitem{CorralOlmedo} D. del-Corral and J. Olmedo, Breaking of isospectrality of quasinormal modes in nonrotating loop quantum gravity black holes, Phys. Rev. D {\bf 105}, 064053 (2022).


\bibitem{Bouh} M. Bouhmadi-L\'opez, S. Brahma, C.-Y. Chen, P. Chen, and D.-h. Yeom, A consistent model of non-singular Schwarzschild black hole in loop quantum gravity and its quasinormal modes, JCAP {\bf 07}, 066 (2020).

\bibitem{QNMloop} C. Zhang and A. Wang, Quasi-normal modes of loop quantum black holes formed from gravitational collapse, JCAP {\bf 10}, 070 (2024).

\bibitem{QNMloop1} L.-G. Zhu, G. Fu, S. Li, D. Zhang, J.-P. Wu, and J.-P. Wu, Quasinormal modes of a charged loop quantum black hole, Phys. Rev. D {\bf 111}, 104008 (2025).

\bibitem{QNMloop2} Z. Dong, S. Li, D. Zhang, and J.-P. Wu, Quasinormal modes of a rotating loop quantum black hole, arXiv:2605.18249.

\bibitem{QNMloop3} E.R. Livine, C. Montagnon, N. Oshita, and H. Roussille, Scalar quasi-normal modes of a loop quantum black hole, JCAP {\bf 10}, 037 (2024).




\bibitem{LitBH1} L. Modesto, Loop quantum black hole, Class. Quantum Grav. {\bf 23}, 5587 (2006).

\bibitem{LitBH2} D.W. Chiou, Phenomenological loop quantum geometry of the Schwarzschild black hole, Phys. Rev. D \textbf{78}, 064040 (2008).


\bibitem{CorichiSingh} A. Corichi and P. Singh, Loop quantum dynamics of Schwarzschild interior revisited, Class. Quantum Grav. {\bf 33}, 055006 (2016).


\bibitem{LitBH3} J. Cortez, W. Cuervo, H.A. Morales-T\'ecotl, and J.C. Ruelas, On effective loop quantum geometry of Schwarzschild interior, Phys. Rev. D {\bf 95}, 064041 (2017).

\bibitem{LitBH4} N. Bodendorfer, F.M. Mele, and J. M\"unch, Effective quantum extended spacetime of polymer Schwarzschild black hole, Class. Quantum Grav. {\bf 36}, 195015 (2019).






\bibitem{AOS} A. Ashtekar, J. Olmedo, and P. Singh, Quantum transfiguration of Kruskal black holes, Phys. Rev. Lett. {\bf 121}, 241301 (2018).

\bibitem{AOS2} A. Ashtekar, J. Olmedo, and P. Singh, Quantum extension of the Kruskal spacetime, Phys. Rev. D {\bf 98}, 126003 (2018).

\bibitem{Properties} A. Ashtekar and J. Olmedo,
Properties of a recent quantum extension of the Kruskal geometry, Int. J. Mod. Phys. D {\bf 29}, 2050076 (2020).

\bibitem{Axial} G.A. Mena Marugán and A. Mínguez-Sánchez, Axial perturbations in Kantowski-Sachs spacetimes and hybrid quantum
cosmology, Phys. Rev. D {\bf 109}, 106009 (2024).

\bibitem{Polar} G.A. Mena Marugán and A. Mínguez-Sánchez, Polar perturbations in Kantowski-Sachs spacetimes and hybrid quantum
cosmology, Phys. Rev. D {\bf 111}, 086024 (2025).

\bibitem{GnR} J. Cortez, B. Elizaga Navascu\'es, G.A. Mena Marug\'an, A. Torres-Caballeros, and J.M. Velhinho, Fock quantization of a Klein-Gordon field in the interior geometry of a nonrotating black hole, Mathematics {\bf 11}, 3922 (2023).

\bibitem{Beatles} J. Cortez, G.A. Mena Marug\'an, A. Torres-Caballeros, and J.M. Velhinho, Time-dependent scalings and Fock quantization of a massless scalar field in Kantowski-Sachs spacetime, Class. Quantum Grav. {\bf 41}, 195002 (2024).

%% Dressed metric approach

\bibitem{dressed1} I. Agullo, A. Ashtekar, and W. Nelson, Quantum gravity extension of the inflationary scenario, Phys. Rev. Lett. \textbf{109}, 251301 (2012).

\bibitem{dressed2} I. Agullo, A. Ashtekar, and W. Nelson, Extension of the quantum theory of cosmological perturbations to the Planck era, Phys. Rev. D {\bf 87}, 043507 (2013).

\bibitem{dressed3} I. Agullo, A. Ashtekar, and W. Nelson, The pre-inflationary dynamics of loop quantum cosmology: Confronting quantum gravity with observations, Class. Quantum Grav. {\bf 30}, 085014 (2013).

\bibitem{dressed4} I. Agullo, W. Nelson, and A. Ashtekar, Preferred instantaneous vacuum for linear scalar fields in cosmological spacetimes, Phys. Rev. D {\bf 91}, 064051 (2015).

%% Hybrid approach

\bibitem{hybrid1} M. Fernández-Méndez, G.A. Mena Marugán, and J. Olmedo, Hybrid quantization of an inflationary universe, Phys. Rev. D {\bf 86}, 024003 (2012).

\bibitem{hybrid2} L. Castelló Gomar, M. Martín-Benito, and G.A. Mena Marugán, Gauge-invariant perturbations in hybrid quantum cosmology, JCAP {\bf 06}, 045 (2015).

\bibitem{hybrid5} F. Benítez Martínez and J. Olmedo, Primordial tensor modes of the early Universe, Phys. Rev. D {\bf 93}, 124008 (2016).

\bibitem{EliMenRev} B. Elizaga Navascu{\'e}s and G. A. Mena Marug{\'a}n, Hybrid loop quantum cosmology: An overview, Front. Astron. Space Sci. {\bf 8}, 81 (2021).

%% Perturbations AOS hybrid

\bibitem{Masterfun} M. Lenzi, G.A. Mena Marug{\'a}n, and A. M{\'\i}nguez-S{\'a}nchez, Master functions and hybrid quantization of perturbed nonrotating black hole interiors, Class. Quantum Grav. {\bf 43}, 105007 (2026).

\bibitem{BlackPumas} B. Elizaga Navascu{\'e}s and A. Torres-Caballeros, Effective Regge--Wheeler equations of a hybrid loop quantum black hole, Phys. Rev. D {\bf 113}, 124031 (2026).

%%

\bibitem{KZZ} R.A. Konoplya, A. Zhidenko, and A.F. Zinhailo, Higher order WKB formula for quasinormal modes and greybody factors: Recipes for quick and accurate calculations, Class. Quantum Grav. {\bf 36}, 155002 (2019).

%%


\bibitem{loopquantBH} B. Elizaga Navascu\'es, G.A. Mena Marug\'an, and A. M\'ingez-S\'anchez, Extended phase space quantization of a black hole interior model in loop quantum cosmology, Phys. Rev. D {\bf 108}, 106001 (2023).



\bibitem{ENGQMM} B. Elizaga Navascu\'es, A. Garc\'{\i}a-Quismondo, and G.A. Mena Marug\'an, 
Space of solutions of the Ashtekar-Olmedo-Singh effective black hole model, Phys. Rev. D {\bf 106}, 063516 (2022).

\bibitem{Immirzi} G. Immirzi, Real and complex connections for canonical gravity, Class. Quantum Grav. {\bf 14}, L177 (1997).

\bibitem{Articulin} A. Garc\'{\i}a-Quismondo and G.A. Mena Marug\'an, Exploring alternatives to the Hamiltonian calculation of the Ashtekar-Olmedo-Singh black hole solution, Front. Astron. Space Sci. {\bf 8}, 701723 (2021).


%%


\bibitem{Chandra2} S. Chandrasekar, On one-dimensional potential barriers having equal reflexion and transmission coefficients, Proc. Roy. Soc. Lond. A {\bf 369}, 425 (1980).


\bibitem{GA2} M. Lenzi, G.A. Mena Marug{\'a}n, A. M{\'\i}nguez-S{\'a}nchez, and C. F. Sopuerta, Hamiltonian formalism, master functions and Darboux transformations for perturbed (interiors and exteriors of) nonrotating black holes, Gen. Relativ. Gravit. {\bf 58}, 56 (2026).


\bibitem{Darboux} G. Darboux, Sur une proposition relative aux \'equations lin\'eaires, C. R. Acad. Sci. (Paris) {\bf 94}, 1456 (1882).

\bibitem{CM1} M. Lenzi and C.F. Sopuerta, Darboux covariance: A hidden symmetry of perturbed Schwarzschild black holes, Phys. Rev. D {\bf 104}, 124068 (2021).

\bibitem{CM2} M. Lenzi and C.F. Sopuerta, Master functions and equations for perturbations of vacuum spherically symmetric spacetimes. Phys. Rev. D {\bf 104}, 084053 (2021).


\bibitem{Frolov} V. Frolov and I. Novikov, Black Hole Physics: Basic Concepts and New Developments (Springer, Dordrecht, 1998).


\bibitem{ShutzWill} B.F. Schutz and C.M. Will, Black hole normal modes: A semianalytic approach, Astrophys. J. Lett. {\bf 291}, L33 (1985).

\bibitem{IyerWill} S. Iyer and C.M. Will, Black hole normal modes: A WKB approach. 1. Foundations and applications of a higher order WKB analysis of potential barrier scattering, Phys. Rev. D {\bf 35}, 3621 (1987).

\bibitem{Opala} J. Matyjasek and M. Opala, Quasinormal modes of black holes. The improved semianalytic approach, Phys. Rev. D {\bf 96}, 024011 (2017).

%%

\bibitem{numericalcode} The Mathematica package implementing the thirteenth WKB order with Padé approximants for the computation of quasinormal modes and greybody factors, together with example calculations for the Schwarzschild black hole, is publicly available at https://goo.gl/nykYGL.

\end{thebibliography}
\end{document}